\documentclass[aps,prd,showkeys,superscriptaddress,nofootinbib,floatfix]{revtex4-1}
\pdfoutput=1 % THIS IS NEEDED TO SUBMIT TO JHEP, OTHERWISE THEIR SYSTEM CANNOT COMPILE THE FILE WITH pdf/jpg/png FIGURES!

\usepackage{float} % use the command [H] after \begin{figure} to FORCE LaTeX to put the figure where you want!
\usepackage{subfiles}
\usepackage{amsmath}
\usepackage{amssymb}
\usepackage{amsthm}
\usepackage{mathrsfs}
\usepackage{graphicx}
\usepackage{epstopdf}
\usepackage{fancyhdr}
\usepackage{array}
\usepackage[all]{xy}
\usepackage{eufrak}
\usepackage{euscript}
\usepackage{enumerate}
\usepackage{slashed}
\usepackage{hyperref}
\usepackage{subfigure} % NEEDED IN ORDER TO USE SUBFIG
\usepackage{epstopdf} % COMPILE EPS (AND PDF) FIGURES USING PDFLATEX!!!
\hypersetup{pdftex,colorlinks=true,linkcolor=blue,citecolor=blue,menucolor=black,urlcolor=blue,filecolor=blue}

\def \mw {\mathfrak{w}}

\begin{document}

\title{Diagonal and Hall holographic conductivities dual to a bulk condensate of magnetic monopoles}

\author{Romulo Rougemont}
\email{romulo.pereira@uerj.br, analisadorcetico at gmail dot com}
\affiliation{Departamento de F\'{i}sica Te\'{o}rica, Universidade do Estado do Rio de Janeiro,
Rua S\~{a}o Francisco Xavier 524, 20550-013, Maracan\~{a}, Rio de Janeiro, Rio de Janeiro, Brazil}

%\date{\today}

\begin{abstract}
By employing the holographic operator mixing technique to deal with coupled perturbations in the gauge/gravity duality, I numerically compute the real and imaginary parts of the diagonal and Hall AC conductivities in a strongly coupled quantum field theory dual to a bulk condensate of magnetic monopoles. The results obtained show that a conclusion previously derived in the literature, namely, the vanishing of holographic DC conductivities in 3-dimensional strongly coupled quantum field theories dual to a 4-dimensional bulk magnetic monopole condensate, also applies to the calculation of diagonal and Hall conductivities in the presence of a topological $\theta$-term. Therefore, the condensation of magnetic monopoles in the bulk is suggested as a rather general and robust mechanism to generate dual strongly coupled quantum field theories with zero DC conductivities. The interplay between frequency, $\theta$-angle and the characteristic mass scale of the monopole condensate on the results for the conductivities is also investigated.
\end{abstract}

%\pacs{Valid PACS appear here}
% PACS, the Physics and Astronomy Classification Scheme.
% Valid PACS numbers may be entered using the \verb+\pacs{#1} command.

% Use showkeys class option if keyword display desired

\maketitle
\tableofcontents

\section{Introduction}
\label{sec:intro}

The role played by magnetic monopoles in the determination of holographic phases of matter have recently attracted attention in the literature. For instance, in Refs. \cite{faulkner-iqbal,sachdev-cond,horowitz-iqbal,iqbal,Alho:2016gdf} the correlations of monopole operators have been investigated in holographic models at finite temperature and density. In particular, the condensation of magnetic monopoles in the bulk, corresponding to the establishment of a bulk dual superconducting phase -- responsible for confining electric fields within flux tubes in the bulk, which pop up as localized electric charges at the boundary of the bulk geometry --, has been suggested as a holographic dual of insulating states of the corresponding strongly coupled quantum field theory (QFT) living at the boundary.

Regarding the calculation of transport coefficients, in Ref. \cite{Rougemont:2015gia} it has been shown that the diagonal holographic DC conductivity in 3-dimensional strongly coupled QFT's at finite temperature -- in the absence of a topological $\theta$-term in the 4-dimensional bulk -- vanishes when the bulk comprises a condensate of magnetic monopoles (defined in the probe limit). This result was analytically demonstrated for a general isotropic background black hole metric at finite temperature. The main question investigated in the present work regards the fate of this result when a topological $\theta$-term is included in the bulk effective action.

In the presence of a $\theta$-term, pure Maxwell theory in the bulk (which corresponds to a bulk phase with no magnetic monopoles) implies nonzero, but frequency-independent values for the diagonal and Hall (off-diagonal) conductivities, as shown in Ref. \cite{Iqbal:2008by}. As I am going to show in the present work, this result is drastically modified when a magnetic monopole condensate is formed within the bulk.

As done in Ref. \cite{Rougemont:2015gia}, the low energy effective action describing the condensate of magnetic monopoles in the bulk will be constructed here by means of the so-called Julia-Toulouse mechanism (JTM), which was originally proposed in Ref. \cite{jt} in the context of non-relativistic condensed matter media as a prescription to identify the lowest lying modes of a system in a phase characterized by a condensate of topological defects. This was later generalized in Refs. \cite{qt,artigao} to deal with the construction of low energy effective actions for relativistic systems described by $p$-forms (corresponding to antisymmetric tensor fields of rank $p$) non-minimally coupled to Dirac-like defects \cite{dirac1,dirac2}, in the regime where these defects proliferate until forming a macroscopically continuous distribution corresponding to the condensate of defects. The JTM has been then applied to describe different aspects of several physical systems, see for instance Refs. \cite{artigao,santiago,mcsmon,dafdc,jt-cho-su2,jt-cho-su3,mbf-m,jt-schwinger,proca-m,jt-localizacao,jt-lorentz,Braga:2016atp}.

Particularly, the $p$-form of interest in the present work is the Maxwell gauge field $A_\mu$ defined in an asymptotically AdS$_4$ background, and the associated defects that couple non-minimally to this 1-form are Dirac-like magnetic monopoles. In Ref. \cite{Rougemont:2015gia} it was discussed in details how the JTM can be implemented in the context of the gauge/gravity duality \cite{adscft1,adscft2,adscft3} to describe the effects of a 4-dimensional bulk condensate of magnetic monopoles in the conductivity of the dual strongly coupled QFT in 3 dimensions. In the present work I generalize this approach to include the effects associated to a topological $\theta$-term in the bulk. In the absence of such term the Hall conductivity trivially vanishes, therefore, one of the main motivations for considering the contribution of the $\theta$-term is to obtain nontrivial results for the Hall conductivity, besides the diagonal conductivity which is also affected by the presence of the $\theta$-term in the magnetically condensed phase of the bulk, as I am going to show.

This work is organized as follows. I begin in section \ref{sec:diluted} by reviewing in details the calculation of the Hall and diagonal conductivities in the probe Maxwell theory with the $\theta$-term defined on top of the AdS$_4$-Schwarzschild background, which in view of the JTM corresponds to a regime of the bulk with no magnetic monopoles (completely diluted phase). The main original results of this work are presented in section \ref{sec:magcond}, where I consider a complete monopole condensation in the bulk and study the corresponding effects on the diagonal and Hall conductivities of the dual boundary QFT. As it will be shown, the main result derived in Ref. \cite{Rougemont:2015gia}, namely, the vanishing of the diagonal DC conductivity in the magnetically condensed phase (in the absence of a $\theta$-term in the bulk) remains valid upon the inclusion of the $\theta$-term, and also holds for the DC Hall conductivity. In this way, one of the main results of the present work is the indication that a bulk monopole condensate constitutes a fairly general and robust holographic mechanism to generate zero DC conductivities in 3-dimensional strongly coupled QFT's. I will also investigate the interplay between the frequency, the $\theta$-angle, and the characteristic mass scale of the monopole condensate on the diagonal and Hall AC conductivities. The numerical results for the AC conductivities in the magnetically condensed phase are obtained through the use of the holographic operator mixing technique \cite{Amado:2009ts,Kaminski:2009dh,Kim:2014bza,Critelli:2016ley}, which is required since the $\theta$-term couples the relevant fluctuations in this phase. Furthermore, in appendix \ref{sec:eltcond} I also investigate the case with (no monopoles and) a complete electric charge condensation in the bulk and the corresponding effects on the boundary QFT diagonal and Hall conductivities.

In this work I use natural units with dimensionless $c=\hbar=k_B=1$ and a mostly plus metric signature. Greek indexes run over all bulk coordinates, while Latin indexes denote only coordinates parallel to the boundary.

\section{Hall and diagonal conductivities in the bulk diluted phase}
\label{sec:diluted}

The purpose of this review section is twofold. First, the calculations reviewed here are completely analytical and serve as a standard and clear example of how to calculate thermal retarded Green's functions in holography for decoupled perturbations \cite{Son:2002sd}. This will be contrasted with the more involved situation discussed in section \ref{sec:magcond}, where one needs to obtain retarded propagators for coupled perturbations in a numerical calculation where the identification of the proper contributions for the diagonal and Hall conductivities is not so straightforward due to the coupled equations of motion for the relevant perturbations. This will be done by means of the more general holographic operator mixing technique \cite{Amado:2009ts,Kaminski:2009dh,Kim:2014bza,Critelli:2016ley}, which agrees with the simpler prescription of \cite{Son:2002sd} when the latter is applicable (as it is the case for the Maxwell theory with a $\theta$-term \cite{Iqbal:2008by} reviewed in the present section, and also for the case with a bulk monopole condensate in the absence of a $\theta$-term, which was first discussed in Ref. \cite{Rougemont:2015gia}). And second, the effective action for the Maxwell theory with a $\theta$-term employed below, which describes the bulk phase without any magnetic monopoles (completely diluted phase), is also the starting point for the construction of the low energy effective action for the bulk monopole condensate via JTM, to be discussed in section \ref{sec:magcond}. Moreover, the analytical results reviewed in the present section will be also important in section \ref{sec:magcond} since, as I will discuss, in the ultraviolet limit of high frequencies one needs to recover the analytical Maxwell results for the diagonal and Hall conductivities in the magnetically condensed phase (this will serve as an important check of the numerical robustness of the method employed in section \ref{sec:magcond}).

I begin working with a general background isotropic metric $g_{\mu\nu}=\textrm{diag}\left(g_{uu},-g_{tt},g_{xx},g_{yy}=g_{xx}\right)$ with a holographic radial coordinate $u$ in terms of which the boundary locates at $u=\epsilon\to 0$ and the background black hole horizon lies at $u=u_H$. The probe Maxwell action with a topological $\theta$-term is given by,\footnote{I absorb the dimensionless gauge coupling constant in $4D$ into the definition of the Maxwell field and take the $\theta$-angle to be independent of the radial holographic coordinate $u$. Notice also that $\tilde{F}^{\mu\nu}=\varepsilon^{\mu\nu\alpha\beta}F_{\alpha\beta}/2\sqrt{-g}$.}
%Notice also that in the second line the constant factor in front of the $\theta$-term is $-1/8$ instead of $1/4$ as written in Eq. (52) of \cite{Iqbal:2008by}; I think that, if the Hall conductivity is to be equals $\theta$ for the Maxwell theory as in Eq. (54) of \cite{Iqbal:2008by}, then the correct factor is $-1/8$, as will be shown here.
\begin{align}
S_{\textrm{dil}}^{\theta}[A_\mu]&=-\frac{1}{4}\int_{\mathcal{M}_4}d^4x\sqrt{-g}F_{\mu\nu}^2 -\frac{\theta}{4}\int_{\mathcal{M}_4}d^4x\sqrt{-g} F_{\mu\nu}\tilde{F}^{\mu\nu}\nonumber\\
&=-\frac{1}{2}\int_{\mathcal{M}_4}d^4x\left[\partial_\mu\left(\sqrt{-g}g^{\mu\alpha}g^{\nu\beta} A_\nu F_{\alpha\beta}\right)- A_\nu \partial_\mu\left(\sqrt{-g}g^{\mu\alpha}g^{\nu\beta} F_{\alpha\beta}\right)\right] -\frac{\theta}{8}\int_{\mathcal{M}_4}d^4x \varepsilon^{\mu\nu\alpha\beta}F_{\mu\nu}F_{\alpha\beta}\nonumber\\
&=-\frac{1}{2}\int_{\mathcal{M}_4}d^4x \partial_\mu\left(\sqrt{-g}g^{\mu\alpha}g^{\nu\beta} A_\nu F_{\alpha\beta} +\theta\varepsilon^{\mu\nu\alpha\beta}A_\nu\partial_\alpha A_\beta\right) +\frac{1}{2}\int_{\mathcal{M}_4}d^4x A_\nu \partial_\mu\left(\sqrt{-g}g^{\mu\alpha}g^{\nu\beta} F_{\alpha\beta}\right)\nonumber\\
&=-\frac{1}{2}\int_{\partial\mathcal{M}_4}d^3x \left(\sqrt{-g}g^{uu}g^{\nu\beta} A_\nu F_{u\beta} +\theta\varepsilon^{u\nu\alpha\beta}A_\nu\partial_\alpha A_\beta\right)\Biggr|_{u=\epsilon}^{u=u_H} +\frac{1}{2}\int_{\mathcal{M}_4}d^4x A_\nu \partial_\mu\left(\sqrt{-g}g^{\mu\alpha}g^{\nu\beta} F_{\alpha\beta}\right),
\label{3.3}
\end{align}
where in the last line I used Stoke's theorem to integrate in the radial direction $u$ and get the border terms. By varying the bulk piece of Eq. \eqref{3.3} with respect to the Maxwell field one obtains the Maxwell equations,\footnote{Notice that the $4D$ equation of motion in the diluted phase is not affected by the presence of the $\theta$-term, since the latter corresponds to a Chern-Simons action at the $3D$ boundary. However, in the magnetically condensed phase the $\theta$-term will also affect the equations of motion and dynamically mix the components of the relevant fields, making the analysis of the problem considerably more involved.}
\begin{align}
\partial_\mu\left(\sqrt{-g}g^{\mu\alpha}g^{\nu\beta} F_{\alpha\beta}\right) = 0,
\label{3.9}
\end{align}
and one sees that the last term in Eq. \eqref{3.3} vanishes on-shell. By following the prescription originally put forward in \cite{Son:2002sd} to obtain the thermal retarded propagator of the vector current sourced by the boundary value of the bulk Maxwell field, one discards the border term located at the horizon in the action \eqref{3.3} and evaluate on-shell the remainder of the action with infalling wave condition for the Maxwell field at the horizon,\footnote{I work here in the radial gauge defined by the condition $A_u=0$. I take the $tt$-component of the metric to be $-g_{tt}$ with $g_{tt}>0$, and use that $\varepsilon^{u\nu\alpha\beta}\equiv\varepsilon^{ijk}\delta_i^\nu\delta_j^\alpha\delta_k^\beta$ with $\varepsilon^{txy}\equiv 1$. Notice also that, since I am considering an isotropic metric, $g_{yy}=g_{xx}$.}
\begin{align}
S_{\textrm{dil}}^{\theta,\textrm{bdy}}[A^0_i]&=+\frac{1}{2}\int_{\partial\mathcal{M}_4}d^3x \lim_{u=\epsilon\to 0}\left\{ \sqrt{-g} g^{uu} \left[-g^{tt}A_tF_{ut} +g^{xx}\left(A_xF_{ux}+A_yF_{uy}\right)\right] +\theta\varepsilon^{ijk}A_i\partial_jA_k\right\} \Biggr|_{\textrm{on-shell}}^{\textrm{infalling}}\nonumber\\
&=\frac{1}{2}\int_{\partial\mathcal{M}_4}d^3x\lim_{u=\epsilon\to 0}\left\{ \sqrt{-g} g^{uu} \left[-g^{tt}A_tA_t' +g^{xx}\left(A_xA_x'+A_yA_y'\right)\right] -\theta\left(A_x\partial_tA_y-A_y\partial_tA_x\right)\right\} \Biggr|_{\textrm{on-shell}}^{\textrm{infalling}},
\label{3.4}
\end{align}
where $A_i^0$ is the boundary value of the Maxwell field and in the last line I considered the Maxwell field as function of just $u$ and $t$ (the prime denotes derivative with respect to $u$), since for the calculation of the electric conductivity one just needs to consider the retarded propagator of the boundary vector current evaluated at zero spatial momentum. I define the Fourier modes $A_i(u,\omega)$ in momentum space according to,\footnote{Since I am omitting the dependence on the spatial momentum, because I will take it to zero in the evaluation of the 2-point retarded correlation function, from now on I will also omit the integration in the spatial directions.}
\begin{align}
A_i(u,t)=\int \frac{d\omega}{2\pi}e^{-i\omega t}A_i(u,\omega).
\label{3.5}
\end{align}
The Dirichlet boundary condition for the Fourier modes reads,
\begin{align}
\lim_{u\rightarrow 0} A_i(u,\omega)=A_i^0(\omega).
\label{3.6}
\end{align}
By substituting \eqref{3.5} into \eqref{3.4}, one gets,
\begin{align}
S_{\textrm{dil}}^{\theta,\textrm{bdy}}[A^0_i]&=\frac{1}{2}\lim_{u=\epsilon\to 0}\int dt \int \frac{d\omega}{2\pi} \int \frac{d\tilde{\omega}}{2\pi} e^{-i(\omega+\tilde{\omega})t} \left\{ \sqrt{-g} g^{uu} \left[-g^{tt}A_t(u,\tilde{\omega}) A_t'(u,\omega) +g^{xx}\left(A_x(u,\tilde{\omega})A_x'(u,\omega)+\right.\right.\right.\nonumber\\
&\left.\left.\left.+ A_y(u,\tilde{\omega})A_y'(u,\omega)\right)\right] +i\omega\theta\left(A_x(u,\tilde{\omega})A_y(u,\omega)-A_y(u,\tilde{\omega})A_x(u,\omega)\right)\right\} \Biggr|_{\textrm{on-shell}}^{\textrm{infalling}}\nonumber\\
&=\frac{1}{2}\lim_{u=\epsilon\to 0}\int \frac{d\omega}{2\pi} \left\{ \sqrt{-g} g^{uu} \left[-g^{tt}A_t(u,-\omega) A_t'(u,\omega) +g^{xx}\left(A_x(u,-\omega)A_x'(u,\omega)+\right.\right.\right.\nonumber\\
&\left.\left.\left.+ A_y(u,-\omega)A_y'(u,\omega)\right)\right] +i\omega\theta\left(A_x(u,-\omega)A_y(u,\omega)-A_y(u,-\omega)A_x(u,\omega)\right)\right\} \Biggr|_{\textrm{on-shell}}^{\textrm{infalling}}.
\label{Smom}
\end{align}
The reality condition for the Maxwell field, $A_\mu^*(u,t)=A_\mu(u,t)$, implies that $A_\mu^*(u,\omega)=A_\mu(u,-\omega)$. Using this result in Eq. \eqref{Smom}, one obtains,
\begin{align}
S_{\textrm{dil}}^{\theta,\textrm{bdy}}[A^0_i]&=-\frac{1}{2}\int \frac{d\omega}{2\pi} \left\{-\lim_{u=\epsilon\to 0}\left( \sqrt{-g} g^{uu} \left[-g^{tt}A_t^*A_t' +g^{xx}\left(A_x^*A_x'+A_y^*A_y'\right)\right] +i\omega\theta\left(A_x^*A_y-A_y^*A_x\right)\right)\right\} \Biggr|_{\textrm{on-shell}}^{\textrm{infalling}}.
\label{Smom2}
\end{align}

By working in the radial gauge, $A_u=0$, in the limit of vanishing spatial momentum, one substitutes the Fourier mode $A_i(u,t,\omega)\equiv e^{-i\omega t}A_i(u,\omega)$ into the Maxwell equations (\ref{3.9}), obtaining the following set of equations of motion for the components of the Maxwell field,
\begin{align}
A_x''+\left(-\frac{g_{uu}'}{g_{uu}}-\frac{g_{xx}'}{g_{xx}}+\frac{\partial_u\sqrt{-g}}{\sqrt{-g}}\right)A_x'
+\frac{g_{uu}\omega^2}{g_{tt}}A_x&=0,\label{3.10}\\
A_y''+\left(-\frac{g_{uu}'}{g_{uu}}-\frac{g_{yy}'}{g_{yy}}+\frac{\partial_u\sqrt{-g}}{\sqrt{-g}}\right)A_y'
+\frac{g_{uu}\omega^2}{g_{tt}}A_y&=0,\label{3.11}\\
A'_t&=0.\label{3.12}
\end{align}
The Dirichlet boundary condition (\ref{3.6}) together with equation (\ref{3.12}) imply that $A_t=A_t^0$. And since I am considering an isotropic metric, $g_{xx}=g_{yy}$ and, therefore, eqs. (\ref{3.10}) and (\ref{3.11}) have the very same structure.

I now specialize to a specific background, given by the metric of the near-horizon approximation of a non-extremal M2-brane solution of $11D$ supergravity, corresponding to a very massive AdS$_4$-Schwarzchild black hole (modulo a 7-sphere),\footnote{See the discussions around Eqs. (119) and (259) of Ref. \cite{Petersen:1999zh}.}
\begin{align}
ds^2=\frac{L^2dU^2}{4U^2f(U)}+\frac{4U^2}{L^2}\left(-f(U)dt^2+dx^2+dy^2\right),
\label{4.1}
\end{align}
where $t,x,y\in(-\infty,\infty)$, $U\in(U_H,\infty)$, $L/2$ is the radius of the asymptotically AdS$_4$ space (corresponding to half the radius $L$ of the 7-sphere \cite{Petersen:1999zh}, which I did not write explicitly above), $f(U)=1-U_H^3/U^3$, $U_H$ is the non-extremality parameter (which vanishes for the extremal solution) and the boundary lies at $U\rightarrow\infty$, with the horizon placed at $U=U_H$. The Hawking temperature of the black hole is then given by,
\begin{align}
T=\frac{\sqrt{g_{tt}'g^{UU}\,'}}{4\pi}\biggr|_{U=U_H}=\frac{3U_H}{\pi L^2}.
\label{3.2}
\end{align}
I now define a new radial coordinate according to,
\begin{align}
u:=\frac{U_H}{U}\Rightarrow f(U)=1-\frac{U_H^3}{U^3}=1-u^3=:h(u),
\label{4.2}
\end{align}
and recast Eq. (\ref{4.1}) in the following form,
\begin{align}
ds^2=\frac{L^2du^2}{4u^2h(u)}+\frac{4U_H^2}{L^2u^2}(-h(u)dt^2+dx^2+dy^2),
\label{4.3}
\end{align}
where in terms of the radial coordinate $u$, the boundary is at $u=0$, while the horizon is placed at $u=1$. By using Eq. (\ref{3.2}), one rewrites Eq. (\ref{4.3}) as follows,
\begin{align}
ds^2=\frac{L^2du^2}{4u^2h(u)}+\frac{4(\pi TL)^2}{9u^2}(-h(u)dt^2+dx^2+dy^2).
\label{4.5}
\end{align}

For the metric (\ref{4.5}), the Maxwell equations (\ref{3.10}), (\ref{3.11}), and (\ref{3.12}), read,
\begin{align}
A_x''+\frac{h'}{h}A_x'+\frac{9\mw^2}{h^2}A_x&=0,\label{4.6}\\
A_y''+\frac{h'}{h}A_y'+\frac{9\mw^2}{h^2}A_y&=0,\label{4.7}\\
A'_t&=0,\label{4.8}
\end{align}
where I defined the dimensionless frequency,
\begin{align}
\mw:=\frac{\omega}{4\pi T}.
\label{4.9}
\end{align}
The set of ordinary differential equations (ODE's) (\ref{4.6}), (\ref{4.7}), and (\ref{4.8}) has analytical solutions, therefore, in the present case, one is able to compute the AC electric conductivity analytically in the bulk diluted phase. As discussed before, one has $A_t=A_t^0$. The general solution of the ODE's (\ref{4.6}) and (\ref{4.7}) can be put in the form below,
\begin{align}
A_{x(y)}(u,\omega)=\frac{C_1+iC_2}{2}\,\tau_+(u)(1-u)^{-i\mw}+\frac{C_1-iC_2}{2}\,\tau_-(u)(1-u)^{i\mw},
\label{4.10}
\end{align}
where $\tau_{\pm}(u)=\exp\left\{\pm\frac{i\mw}{2}\left[2\sqrt{3}\arctan\left(\frac{1+2u}{\sqrt{3}}\right) +\ln(1+u+u^2)\right]\right\}$ are regular functions at the horizon $u=1$. The solution $\propto(1-u)^{-i\mw}$ corresponds to a wave travelling to the horizon and, therefore, the infalling wave condition at the horizon is imposed by setting $C_1=iC_2\equiv C$ in Eq. (\ref{4.10}),
\begin{align}
A_{x(y)}(u,\omega)=C\tau_+(u)(1-u)^{-i\mw}.
\label{4.11}
\end{align}
The constant $C$ is fixed by imposing the Dirichlet boundary condition (\ref{3.6}) into (\ref{4.11}),
\begin{align}
C=A_{x(y)}^0 e^{-\frac{i\pi\mw}{2\sqrt{3}}}=A_{x(y)}^0 e^{-\frac{i\omega}{8\sqrt{3}T}},
\label{4.12} 
\end{align}
therefore,
\begin{align}
A_{x(y)}(u,\omega)=A_{x(y)}^0 e^{-\frac{i\omega}{8\sqrt{3}T}}(1-u)^{-\frac{i\omega}{4\pi T}} \exp\left\{\frac{i\omega}{8\pi T}\left[2\sqrt{3}\arctan\left(\frac{1+2u}{\sqrt{3}}\right) +\ln(1+u+u^2)\right]\right\}.
\label{4.13}
\end{align}
From (\ref{4.13}), one obtains,
\begin{align}
\lim_{u\rightarrow 0} A_{x(y)}(u,\omega)=A_{x(y)}^0(\omega) \qquad \textrm{and} \qquad \lim_{u\rightarrow 0} A'_{x(y)}(u,\omega)=\frac{3i\omega}{4\pi T}A_{x(y)}^0(\omega).
\label{4.14}
\end{align}

For the metric \eqref{4.5}, one has at the boundary,
\begin{align}
\lim_{u\to 0}\sqrt{-g}g^{uu}g^{xx}=\frac{4\pi T}{3}.
\label{asmet}
\end{align}

Substituting Eqs. \eqref{4.8}, \eqref{4.14}, and \eqref{asmet} into the on-shell boundary action \eqref{Smom2}, one obtains,
\begin{align}
S_{\textrm{dil}}^{\theta,\textrm{bdy}}[A^0_i]&=-\frac{1}{2}\int \frac{d\omega}{2\pi} \left[-i\omega\left(A_x^0\,^*(\omega)A_x^0(\omega)+A_y^0\,^*(\omega)A_y^0(\omega)\right) -i\omega\theta\left(A_x^0\,^*(\omega)A_y^0(\omega)-A_y^0\,^*(\omega)A_x^0(\omega)\right)\right],
\label{Smom3}
\end{align}
from which, by following the prescription of Ref. \cite{Son:2002sd}, one extracts the following nontrivial retarded correlators in the bulk diluted phase,
\begin{align}
G_{xx}^{(R),\textrm{dil}}(\omega)=G_{yy}^{(R),\textrm{dil}}(\omega)=-i\omega \qquad \textrm{and} \qquad G_{xy}^{(R),\textrm{dil}}=-G_{yx}^{(R),\textrm{dil}}=-i\omega\theta.
\label{dilprop}
\end{align}
From linear response theory, one has the following Kubo formulas for the diagonal and Hall electric conductivities \cite{Iqbal:2008by},
\begin{align}
\sigma_{xx}^{\textrm{dil}}(\omega)=\sigma_{yy}^{\textrm{dil}}(\omega)=-\frac{G_{xx}^{(R),\textrm{dil}}(\omega)}{i\omega}=1 \qquad \textrm{and} \qquad \sigma_{xy}^{\textrm{dil}}(\omega)=-\sigma_{yx}^{\textrm{dil}}(\omega)=-\frac{G_{xy}^{(R),\textrm{dil}}(\omega)}{i\omega}= \theta.
\label{dilcond}
\end{align}
These results for the diagonal conductivities were first obtained in Refs. \cite{Herzog:2002fn,Herzog:2007ij}, and for the Hall conductivities, in Ref. \cite{Iqbal:2008by}. Since these AC conductivities are frequency-independent, they coincide with the corresponding DC conductivities, given by the zero frequency limit of the AC conductivities.

\section{Hall and diagonal conductivities in the bulk magnetically condensed phase}
\label{sec:magcond}

Now I consider the case with a condensate of magnetic monopoles in the bulk in the presence of a $\theta$-term. The case with no $\theta$-term was originally presented in Ref. \cite{Rougemont:2015gia}, where a vanishing DC conductivity at the boundary QFT was identified as an universal phenomenon dual to a bulk condensate of magnetic monopoles.

By including magnetic defects with charge $\bar{g}$ into the Maxwell strength tensor and considering a complete condensation of the monopoles in the bulk, one obtains through the JTM \cite{Rougemont:2015gia},
\begin{align}
F_{\mu\nu}=\partial_{[\mu}A_{\nu]}-\bar{g}\tilde{\chi}_{\mu\nu} \stackrel{\textrm{cond}}{\longrightarrow} K_{\mu\nu},
\label{jtm}
\end{align}
where $\tilde{\chi}_{\mu\nu}$ is the Chern-Kernel localizing the Dirac brane whose border corresponds to the physical monopole current in the bulk. The condensation of the monopoles leads to the emerging massive 2-form Kalb-Ramond field $K_{\mu\nu}$, where the Chern-Kernel and, therefore, the Kalb-Ramond field, are subjected to the following boundary condition \cite{Rougemont:2015gia},
\begin{align}
\lim_{u\to 0}\tilde{\chi}_{\mu\nu} = 0 \Rightarrow \lim_{u\to 0} K_{\mu\nu}(u,t,\omega)= \lim_{u\to 0}\partial_{[\mu}A_{\nu]}(u,t,\omega) \Rightarrow K_{ij}^0(\omega)= -i\omega A_{x(y)}^0(\omega) \delta^t_{[i}\delta^{x(y)}_{j]},
\label{KRbc}
\end{align}
meaning that the monopole current vanishes at the boundary and the Kalb-Ramond field reduces to the Maxwell strength tensor without defects at the boundary. This boundary condition is chosen in order to have the boundary value of the bulk massive 2-form field $K_{\mu\nu}$ sourcing a vector current operator at the boundary, since it is written in terms of the boundary value of a Maxwell field. The physical interpretation of this picture, detailed discussed in Ref. \cite{Rougemont:2015gia}, is that the conductivity associated to the vector current operator sourced by the boundary value of the Maxwell field can be calculated in different holographic phases. For the bulk diluted phase, the results reviewed in the previous section implied constant and finite conductivities given by Eqs. \eqref{dilcond}. On the other hand, in the bulk magnetically condensed phase without the $\theta$-term, the (diagonal) AC conductivity must be obtained numerically for a given background metric, but for any isotropic black hole metric it can be analytically shown that in the ultraviolet limit of large frequencies the AC conductivity reduces to the same value obtained in Maxwell theory, while in the deep infrared the DC conductivity exactly vanishes \cite{Rougemont:2015gia}.

Now I want to investigate the effects of the $\theta$-term on the diagonal AC conductivity, and also evaluate the Hall (off-diagonal) AC conductivity (which trivially vanishes in the absence of the $\theta$-term) in the bulk magnetically condensed phase. For this sake, as done in Ref. \cite{Rougemont:2015gia}, one substitutes the JTM prescription \eqref{jtm} into the Maxwell action \eqref{3.3}, and in order to complete the construction of a low-energy effective field theory for the magnetically condensed phase, one supplies a dynamics for the emerging massive Kalb-Ramond field $K_{\mu\nu}$ by considering a derivative expansion and retaining only the term with lowest order in derivatives, which gives the dominant contribution at low energies,
\begin{align}
S_{\textrm{cond}}^{\theta}[K_{\mu\nu}]&=-\frac{1}{4}\int_{\mathcal{M}_4}d^4x\left\{\sqrt{-g}\left[\frac{1}{3m^2(u)} F_{\mu\alpha\beta}^2 + K_{\mu\nu}^2\right]+ \frac{\theta}{2} \varepsilon^{\mu\nu\alpha\beta} K_{\mu\nu} K_{\alpha\beta} \right\},
\label{SKR}
\end{align}
where $F_{\mu\alpha\beta}=\partial_\mu K_{\alpha\beta}+\partial_\alpha K_{\beta\mu} +\partial_\beta K_{\mu\alpha}$ is the Kalb-Ramond strength tensor and $m(u)$ is a radial-dependent effective mass for the Kalb-Ramond field. As detailed discussed in Ref. \cite{Rougemont:2015gia}, this radial-dependent mass actually corresponds to a scalar field whose excitations describe the monopoles (which are themselves higher energy excitations in the condensed phase, while the lowest-lying modes in this phase correspond to spin 1 particles associated to the massive Kalb-Ramond field). In the low-energy effective action \eqref{SKR} one neglects the dynamics of this scalar field and takes it as a prescribed profile associated to some specific condensation process in the bulk (which, in turn, corresponds to some specific choice for the potential of this scalar field in the ultraviolet completion of the action \eqref{SKR}, as discussed in Ref. \cite{Rougemont:2015gia}). Notice that the boundary condition \eqref{KRbc} actually follows by imposing that the effective mass $m(u)$ must vanish at the boundary, since the requirement of finiteness of the effective action \eqref{SKR} implies that, in this case, $F_{\mu\alpha\beta}=0$ at the boundary, which is identically satisfied as the Jacobi identity by taking \eqref{KRbc}. Therefore, any prescribed profile one uses for the effective mass in this work will be such that it vanishes at the boundary (while also being regular at the black hole horizon).

Let me now work out the action \eqref{SKR} to identify its border terms and the equation of motion for the massive Kalb-Ramond field in the presence of a $\theta$-term,
\begin{align}
S_{\textrm{cond}}^{\theta}[K_{\mu\nu}]&=-\frac{1}{4}\int_{\mathcal{M}_4}d^4x\left\{\sqrt{-g}\left[\frac{1}{m^2(u)} \partial_\mu K_{\alpha\beta}g^{\mu\rho}g^{\alpha\lambda}g^{\beta\nu}F_{\rho\lambda\nu} + K_{\alpha\beta}g^{\alpha\mu}g^{\beta\nu} K_{\mu\nu}\right]+ \frac{\theta}{2} K_{\alpha\beta} \varepsilon^{\mu\nu\alpha\beta} K_{\mu\nu} \right\}\nonumber\\
&=-\frac{1}{4}\int_{\mathcal{M}_4}d^4x\left\{\partial_\mu\left(\frac{\sqrt{-g}}{m^2(u)} K_{\alpha\beta} g^{\mu\rho}g^{\alpha\lambda}g^{\beta\nu}F_{\rho\lambda\nu}\right)- K_{\alpha\beta} \left[\partial_\mu \left( \frac{\sqrt{-g}}{m^2(u)} g^{\mu\rho}g^{\alpha\lambda}g^{\beta\nu}F_{\rho\lambda\nu} \right) - \left(\sqrt{-g} g^{\alpha\mu}g^{\beta\nu} +\right.\right.\right.\nonumber\\
&\left.\left.\left. + \frac{\theta}{2} \varepsilon^{\mu\nu\alpha\beta}\right) K_{\mu\nu} \right]\right\}\nonumber\\
&=-\frac{1}{4}\int_{\partial\mathcal{M}_4}d^3x\frac{\sqrt{-g}}{m^2(u)} g^{uu}g^{\alpha\lambda}g^{\beta\nu} K_{\alpha\beta} F_{u\lambda\nu}\biggr|_{u=\epsilon}^{u=u_H} +\frac{1}{4}\int_{\mathcal{M}_4}d^4x K_{\alpha\beta} \left[ \partial_\mu \left( \frac{\sqrt{-g}}{m^2(u)} g^{\mu\rho}g^{\alpha\lambda}g^{\beta\nu}F_{\rho\lambda\nu} \right) +\right.\nonumber\\
&\left. - \left(\sqrt{-g} g^{\alpha\mu}g^{\beta\nu} + \frac{\theta}{2} \varepsilon^{\mu\nu\alpha\beta}\right) K_{\mu\nu} \right].
\label{SKR2}
\end{align}
By varying the bulk piece of Eq. \eqref{SKR2}, one gets the equations of motion for the massive Kalb-Ramond field in the presence of the $\theta$-term,
\begin{align}
\partial_\mu \left( \frac{\sqrt{-g}}{m^2(u)} g^{\mu\rho}g^{\alpha\lambda}g^{\beta\nu}F_{\rho\lambda\nu} \right) - \left(\sqrt{-g} g^{\alpha\mu}g^{\beta\nu} + \frac{\theta}{2} \varepsilon^{\mu\nu\alpha\beta}\right) K_{\mu\nu} = 0.
\label{KReom}
\end{align}
As before, by discarding the border piece of the action \eqref{SKR2} evaluated at the horizon, one obtains the following on-shell boundary action,
\begin{align}
S_{\textrm{cond}}^{\theta,\textrm{bdy}}[K_{\mu\nu}^0]&=+\frac{1}{4}\int_{\partial\mathcal{M}_4}d^3x \lim_{u=\epsilon\to 0} \sqrt{-g} g^{uu}g^{\alpha\lambda}g^{\beta\nu} \frac{K_{\alpha\beta}}{m^2(u)} \left( \partial_u K_{\lambda\nu} + \partial_\lambda K_{\nu u} + \partial_\nu K_{u\lambda} \right) \biggr|_{\textrm{on-shell}}^{\textrm{infalling}}.
\label{SKR3}
\end{align}
It is interesting to note that, contrary to what happened in the diluted phase discussed in the previous section, where the $\theta$-term had no effect on the Maxwell equations and contributed directly to the boundary action in the form of a $3D$ Chern-Simons term, in the condensed phase one sees from Eqs. \eqref{KReom} and \eqref{SKR3} that the $\theta$-term modifies the equation of motion for the massive Kalb-Ramond field, while the form of the \emph{off-shell} boundary action is identical to the one obtained in Ref. \cite{Rougemont:2015gia} in the absence of the $\theta$-term; however, the \emph{on-shell} boundary action \eqref{SKR3} does depend on the $\theta$-term through the on-shell values of the components of the Kalb-Ramond field which solve Eq. \eqref{KReom} with infalling wave condition at the black hole horizon.

Let us analyze the components of the equations of motion \eqref{KReom}. There are 3 second order ODE's for 3 independent dynamical variables, $K_{tx}$, $K_{ty}$, and $K_{xy}$ (with the ODE's for $K_{tx}$ and $K_{ty}$ being coupled by the $\theta$-term), and 3 constraints that express $K_{ut}$ as function of $K_{xy}$, $K_{ux}$ as function of $K_{ty}$ and $K'_{tx}$, and $K_{uy}$ as function of $K_{tx}$ and $K'_{ty}$. By substituting the Fourier mode $K_{\mu\nu}(u,t,\omega)\equiv e^{-i\omega t}K_{\mu\nu}(u,\omega)$ into Eq. \eqref{KReom}, one obtains the following set of constraints,\footnote{As before, I recall that the $tt$-component of the metric is $-g_{tt}$ with $g_{tt}>0$, and since I consider an isotropic metric, $g_{yy}=g_{xx}$.}
\begin{align}
K_{ut}&=\theta \frac{\sqrt{g_{tt}g_{uu}}}{g_{xx}} K_{xy},\label{Kuteom}\\
K_{ux}&=\frac{i\omega K'_{tx}-\theta\sqrt{g_{tt}g_{uu}}m^2(u)K_{ty}}{\omega^2-m^2(u)g_{tt}},\label{Kuxeom}\\
K_{uy}&=\frac{i\omega K'_{ty}+\theta\sqrt{g_{tt}g_{uu}}m^2(u)K_{tx}}{\omega^2-m^2(u)g_{tt}},\label{Kuyeom}
\end{align}
while the dynamical equations of motion read as follows,
\begin{align}
0&=K''_{xy}-\left[\frac{g'_{xx}}{g_{xx}}+\frac{g'_{uu}}{2g_{uu}}-\frac{g'_{tt}}{2g_{tt}}+\frac{2m'(u)}{m(u)}\right]K'_{xy} +\frac{g_{uu}}{g_{tt}}\left[\omega^2-m^2(u)g_{tt}(1+\theta^2)\right]K_{xy},\label{Kxyeom}\\
0&=K''_{tx}+\left[\frac{g'_{tt}}{2g_{tt}}-\frac{g'_{uu}}{2g_{uu}}+\frac{m^2(u)g_{tt}}{\omega^2-m^2(u)g_{tt}} \left(\frac{g'_{tt}}{g_{tt}}+\frac{2m'(u)}{m(u)}\right)\right]K'_{tx} +\frac{g_{uu}}{g_{tt}}\left[\omega^2-m^2(u)g_{tt}(1+\theta^2)\right]K_{tx}+\nonumber\\
&+\sqrt{\frac{g_{uu}}{g_{tt}}}\frac{i\omega\theta m(u)\left[ m(u)g'_{tt} +2m'(u) g_{tt} \right]}{\omega^2-m^2(u)g_{tt}}K_{ty},\label{Ktxeom}\\
0&=K''_{ty}+\left[\frac{g'_{tt}}{2g_{tt}}-\frac{g'_{uu}}{2g_{uu}}+\frac{m^2(u)g_{tt}}{\omega^2-m^2(u)g_{tt}} \left(\frac{g'_{tt}}{g_{tt}}+\frac{2m'(u)}{m(u)}\right)\right]K'_{ty} +\frac{g_{uu}}{g_{tt}}\left[\omega^2-m^2(u)g_{tt}(1+\theta^2)\right]K_{ty}+\nonumber\\
&-\sqrt{\frac{g_{uu}}{g_{tt}}}\frac{i\omega\theta m(u)\left[ m(u)g'_{tt} +2m'(u) g_{tt} \right]}{\omega^2-m^2(u)g_{tt}}K_{tx}.\label{Ktyeom}
\end{align}

One sees that the equations of motion \eqref{Ktxeom} and \eqref{Ktyeom} for the variables $K_{tx}$ and $K_{ty}$ are coupled in the presence of the $\theta$-term. One also sees from the above equations that $K_{xy}$ and, therefore, $K_{ut}$, are decoupled from $K_{tx(y)}$, consequently, these components of the Kalb-Ramond field are irrelevant for the calculation of the conductivities. On the other hand, the constraints for $K_{ux(y)}$ must be used to introduce in the on-shell action all the terms depending on $K_{tx(y)}$ and their derivatives. Since $K_{tx(y)}(u=0,\omega)=K_{tx(y)}^0(\omega)=-i\omega A_{x(y)}^0(\omega)$, one needs to collect in the on-shell action all terms proportional to $K_{tx(y)}^{0\,*}K_{tx(y)}^0=\omega^2 A_{x(y)}^{0\,*}A_{x(y)}^0$ in order to obtain the diagonal conductivities $\sigma_{xx}^{\textrm{cond}}(\omega)$ and $\sigma_{yy}^{\textrm{cond}}(\omega)$, while for the calculation of the Hall conductivities $\sigma_{xy}^{\textrm{cond}}(\omega)$ and $\sigma_{yx}^{\textrm{cond}}(\omega)$, one has to collect in the on-shell action all terms proportional to $K_{tx(y)}^{0\,*}K_{ty(x)}^0=\omega^2 A_{x(y)}^{0\,*}A_{y(x)}^0$. Since $K'_{tx(y)}(u=0,\omega)$ may depend both on $K_{tx}^0(\omega)$ and $K_{ty}^0(\omega)$ due to the coupled equations of motion \eqref{Ktxeom} and \eqref{Ktyeom}, which need to be solved numerically, it is not immediately obvious how to disentangle the contributions for the diagonal and Hall conductivities. However, from the above considerations, one may already identify the specific sector of the on-shell boundary action \eqref{SKR3} relevant for the evaluation of these conductivities,\footnote{Since I am going to consider again the propagator evaluated at zero spatial momentum, I already set to zero all the spatial derivatives.}
\begin{align}
S_{\textrm{cond}}^{\theta,\textrm{bdy}}[K_{\mu\nu}^0]&=\frac{1}{4}\int_{\partial\mathcal{M}_4}d^3x \lim_{u=\epsilon\to 0} \frac{\sqrt{-g}g^{uu}}{m^2(u)} \left[2(-g^{tt})g^{xx}K_{tx}\left(K'_{tx}+\partial_t K_{xu}\right) + 2(-g^{tt})g^{yy}K_{ty}\left(K'_{ty}+\partial_t K_{yu}\right) +\right.\nonumber\\
&+\left. 2g^{xx}g^{yy}K_{xy}K'_{xy}\right] \biggr|_{\textrm{on-shell}}^{\textrm{infalling}}\nonumber\\
&=-\frac{1}{2}\int_{\partial\mathcal{M}_4}d^3x \lim_{u=\epsilon\to 0} \frac{1}{\sqrt{g_{uu}g_{tt}}m^2(u)} \left[K_{tx}\left(K'_{tx}-\partial_t K_{ux}\right) + K_{ty}\left(K'_{ty}-\partial_t K_{uy}\right)\right]\biggr|_{\textrm{on-shell}}^{\textrm{infalling}}+\nonumber\\
&+\frac{1}{2}\int_{\partial\mathcal{M}_4}d^3x \lim_{u=\epsilon\to 0} \frac{\sqrt{-g}g^{uu}\left(g^{xx}\right)^2}{m^2(u)} K_{xy}K'_{xy} \biggr|_{\textrm{on-shell}}^{\textrm{infalling}}.
\label{SKR4}
\end{align}
One sees that the last term in the action above is completely decoupled from the relevant sector for the calculation of the conductivities, therefore, one can ignore it in these calculations and work only with the following sector of the on-shell boundary action,\footnote{I already write down the relevant action in momentum space.}
\begin{align}
\mathcal{S}_{\textrm{cond}}^{\theta,\textrm{bdy}}[K_{tx}^0,K_{ty}^0]&=-\frac{1}{2}\int\frac{d\omega}{2\pi}\left\{ \lim_{u=\epsilon\to 0} \frac{1}{\sqrt{g_{uu}g_{tt}}m^2(u)} \left[K_{tx}^*\left(K'_{tx}+i\omega K_{ux}\right) + K_{ty}^*\left(K'_{ty}+i\omega K_{uy}\right)\right]\right\}\biggr|_{\textrm{on-shell}}^{\textrm{infalling}}\nonumber\\
&=-\frac{1}{2}\int\frac{d\omega}{2\pi}\left\{- \lim_{u=\epsilon\to 0} \left[\sqrt{\frac{g_{tt}}{g_{uu}}} \frac{\left(K_{tx}^*K'_{tx}+K_{ty}^*K'_{ty}\right)}{\omega^2-m^2(u)g_{tt}}+ \frac{i\omega\theta\left(K_{tx}^*K_{ty}-K_{ty}^*K_{tx}\right)}{\omega^2-m^2(u)g_{tt}} \right]\biggr|_{\textrm{on-shell}}^{\textrm{infalling}}\right\},
\label{SKR5}
\end{align}
where in the last line I imposed the on-shell constraints \eqref{Kuxeom} and \eqref{Kuyeom}. Notice that for $\theta=0$ the above action reduces to the same one evaluated in Ref. \cite{Rougemont:2015gia}. For $m(u)=0$, since it implies $K_{tx(y)}=\partial_{[t}A_{x(y)]}=-i\omega A_{x(y)}$, one can easily check that the results reviewed in section \ref{sec:diluted} are fully recovered.

Now the final task is to numerically solve the coupled ODE's \eqref{Ktxeom} and \eqref{Ktyeom} for the AdS$_4$-Schwarzschild background \eqref{4.5} with the Dirichlet boundary conditions \eqref{KRbc} and infalling wave conditions for $K_{tx}$ and $K_{ty}$ at the horizon, then substitute these solutions back into the boundary on-shell action \eqref{SKR5}, and finally identify the diagonal and non-diagonal Hall conductivities. This task may be accomplished by using the technique of holographic operator mixing discussed in Refs. \cite{Amado:2009ts,Kaminski:2009dh,Kim:2014bza,Critelli:2016ley}. Below I closely follow the general approach of Ref. \cite{Kim:2014bza} (the interested reader should consult it for a detailed discussion).

\begin{figure}
\begin{tabular}{cc}
\includegraphics[width=0.48\textwidth]{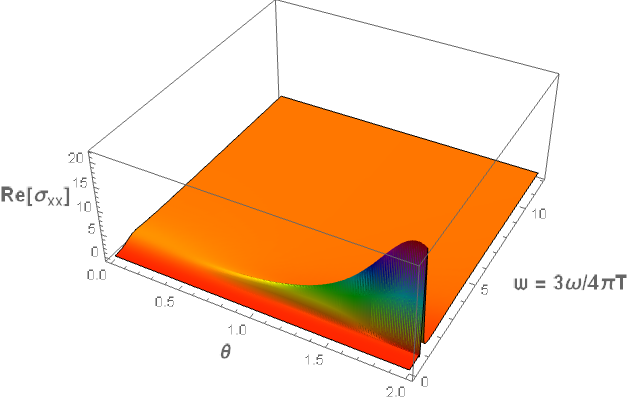} & \includegraphics[width=0.48\textwidth]{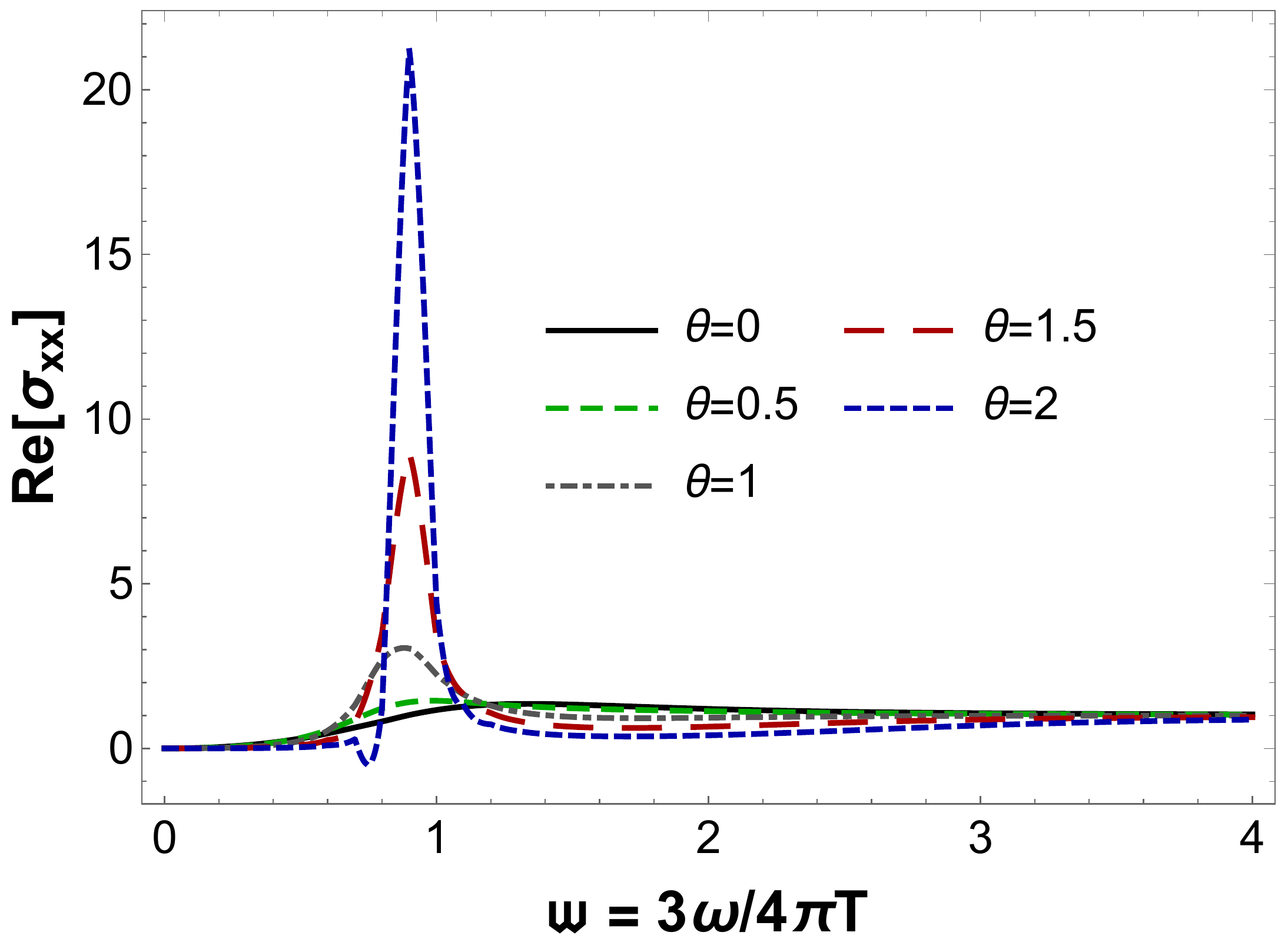}
\end{tabular}
\begin{tabular}{cc}
\includegraphics[width=0.48\textwidth]{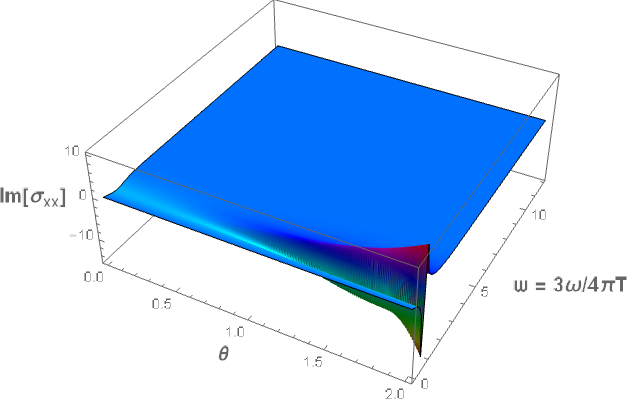} & \includegraphics[width=0.48\textwidth]{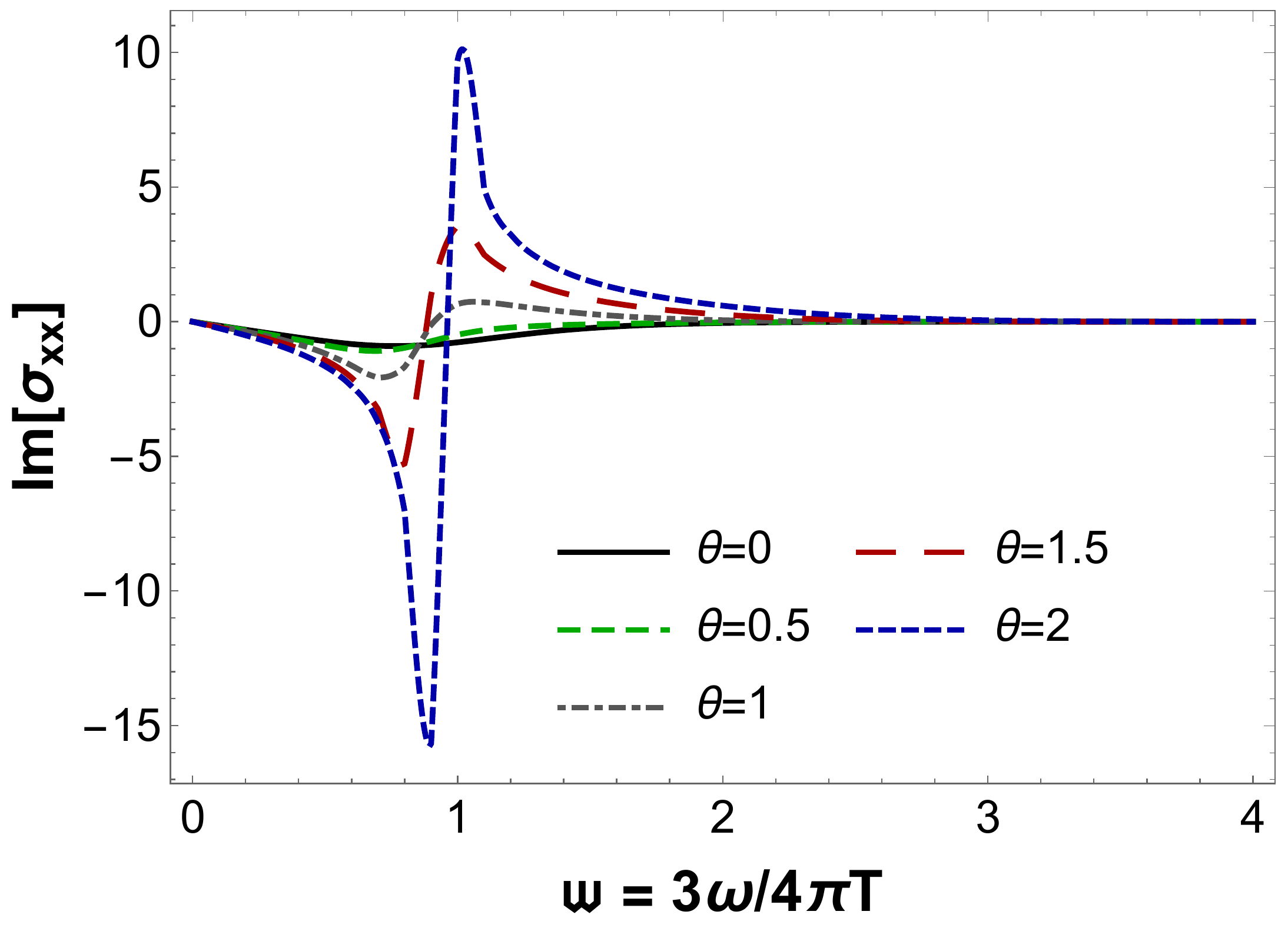}
\end{tabular}
\caption{(Color online) Real and imaginary parts of the diagonal AC conductivity as functions the $\theta$-angle and the dimensionless frequency variable, $\mw=3\omega/4\pi T$, for $C_\Lambda=\Lambda L/2=1$ (dimensionless combination involving the characteristic mass scale $\Lambda$ of the bulk monopole condensate). These results were generated with the mass function $M(u)=\tanh(u)$. \label{fig1}}
\end{figure}

\begin{figure}
\begin{tabular}{cc}
\includegraphics[width=0.48\textwidth]{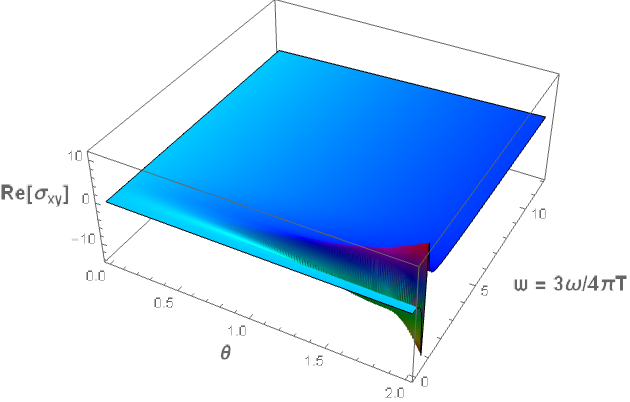} & \includegraphics[width=0.48\textwidth]{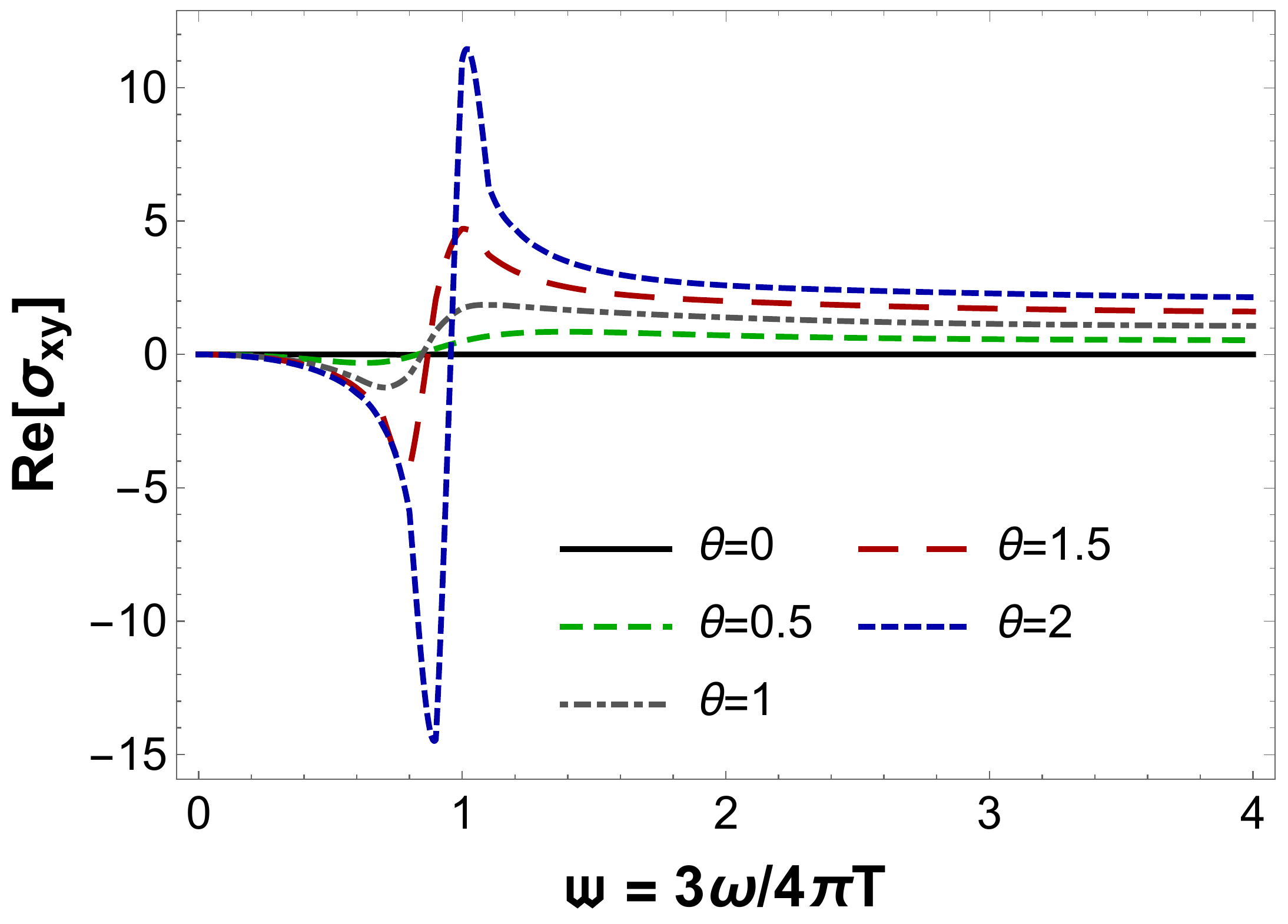}
\end{tabular}
\begin{tabular}{cc}
\includegraphics[width=0.48\textwidth]{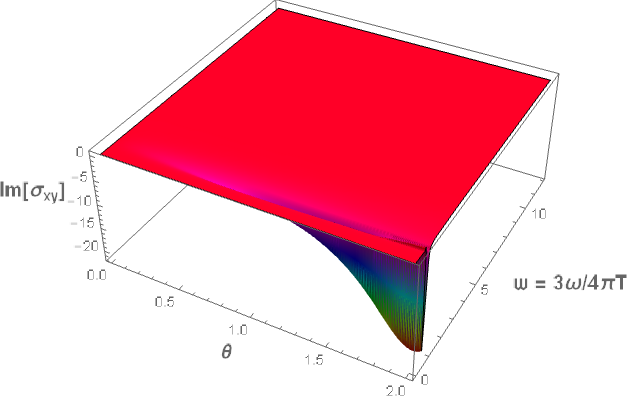} & \includegraphics[width=0.48\textwidth]{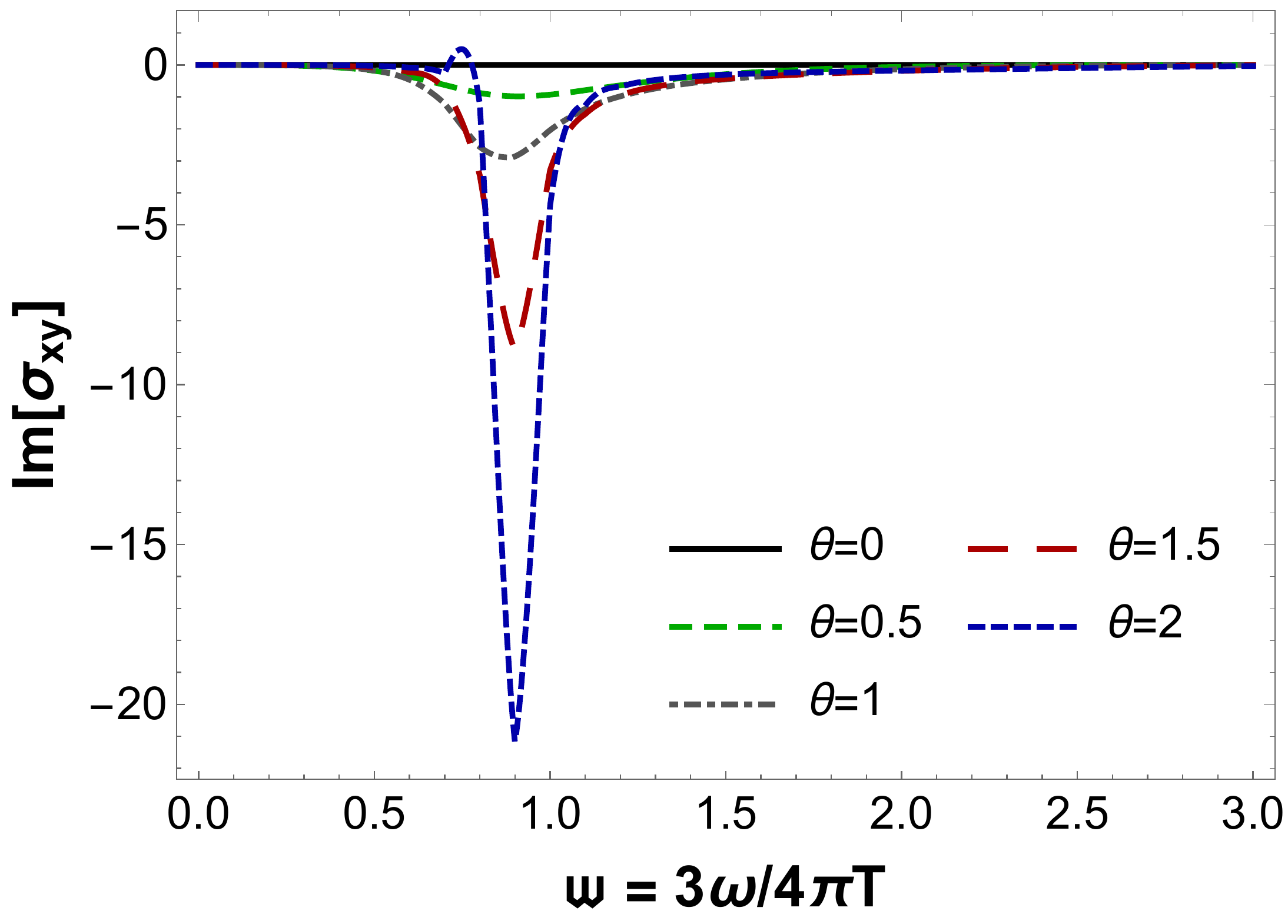}
\end{tabular}
\caption{(Color online) Real and imaginary parts of the Hall AC conductivity as functions the $\theta$-angle and the dimensionless frequency variable, $\mw=3\omega/4\pi T$, for $C_\Lambda=\Lambda L/2=1$ (dimensionless combination involving the characteristic mass scale $\Lambda$ of the bulk monopole condensate). These results were generated with the mass function $M(u)=\tanh(u)$. \label{fig2}}
\end{figure}

\begin{figure}
\begin{tabular}{cc}
\includegraphics[width=0.48\textwidth]{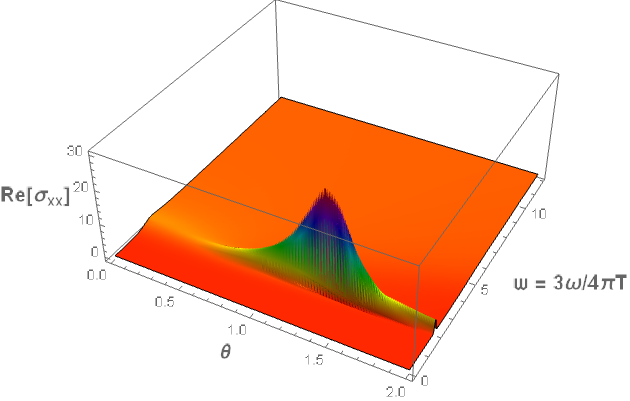} & \includegraphics[width=0.48\textwidth]{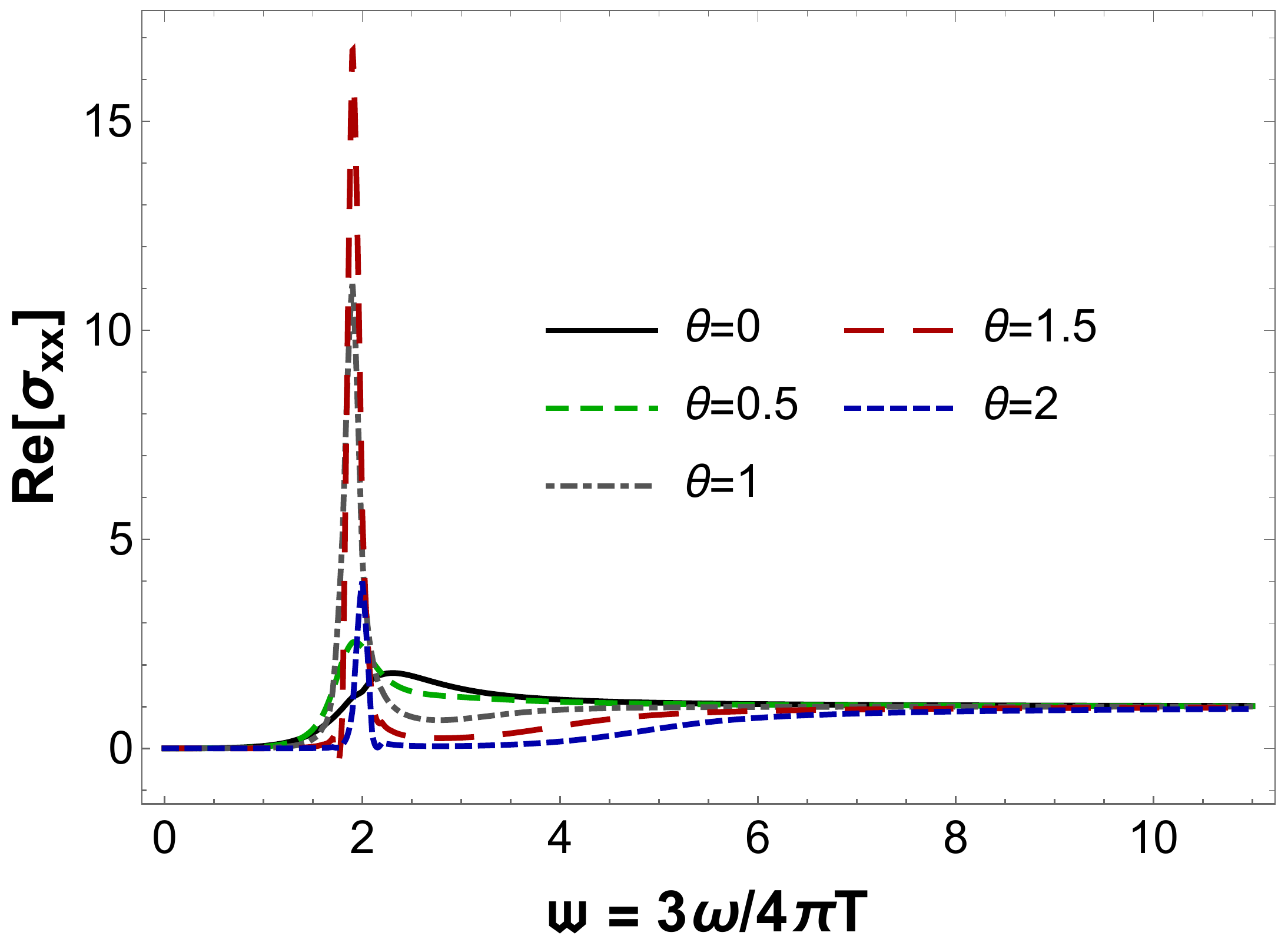}
\end{tabular}
\begin{tabular}{cc}
\includegraphics[width=0.48\textwidth]{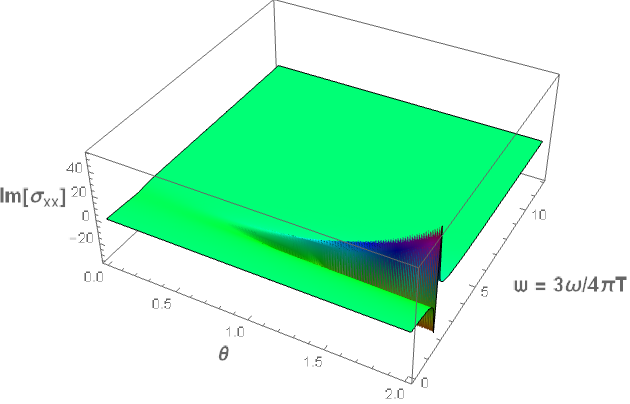} & \includegraphics[width=0.48\textwidth]{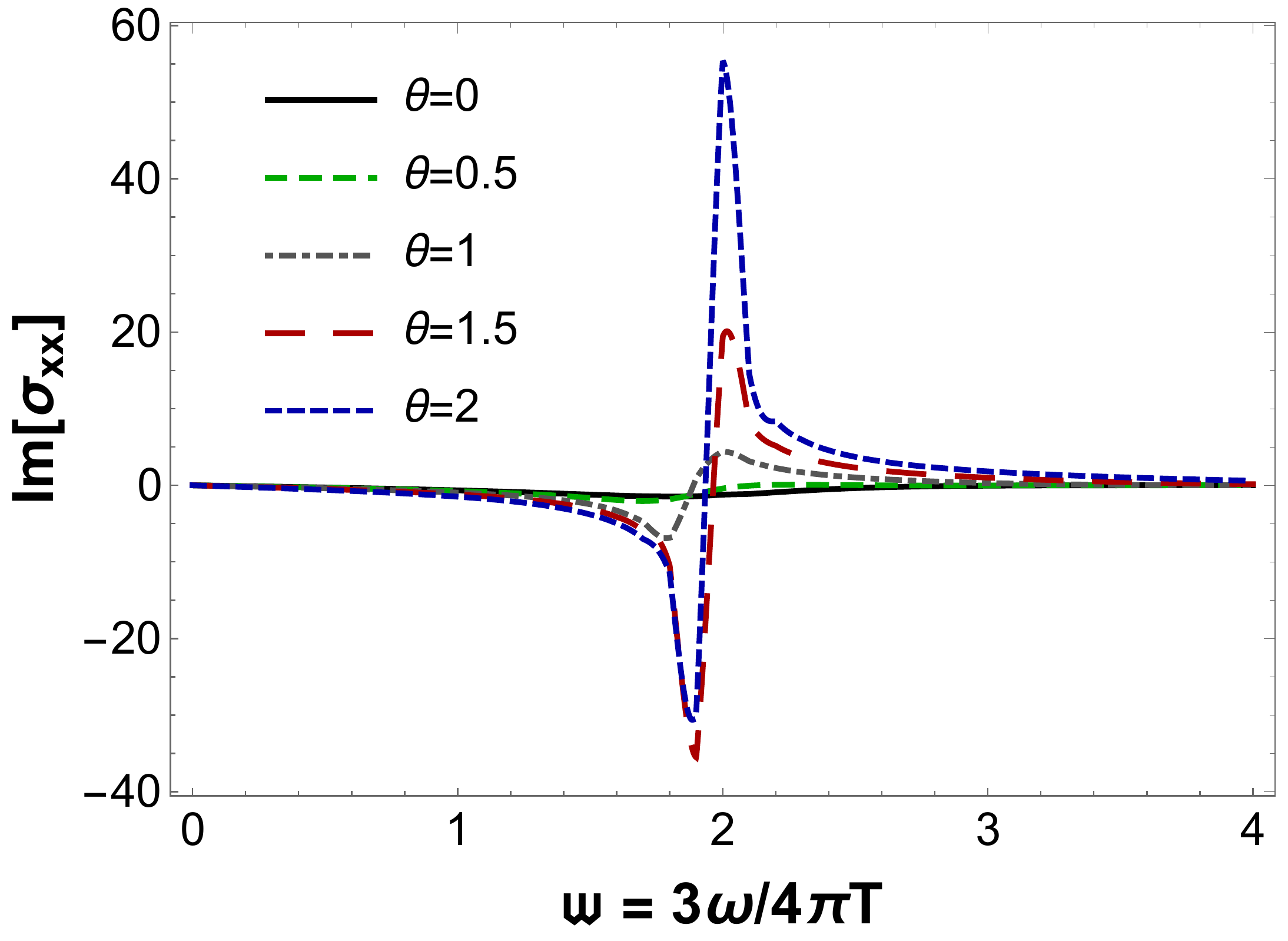}
\end{tabular}
\caption{(Color online) Real and imaginary parts of the diagonal AC conductivity as functions the $\theta$-angle and the dimensionless frequency variable, $\mw=3\omega/4\pi T$, for $C_\Lambda=\Lambda L/2=2$ (dimensionless combination involving the characteristic mass scale $\Lambda$ of the bulk monopole condensate). These results were generated with the mass function $M(u)=\tanh(u)$. \label{fig3}}
\end{figure}

\begin{figure}
\begin{tabular}{cc}
\includegraphics[width=0.48\textwidth]{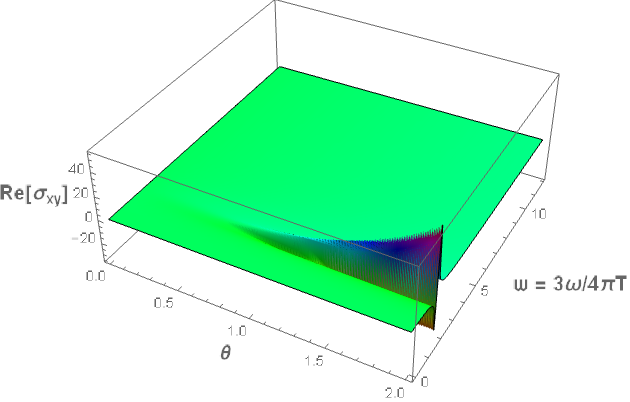} & \includegraphics[width=0.48\textwidth]{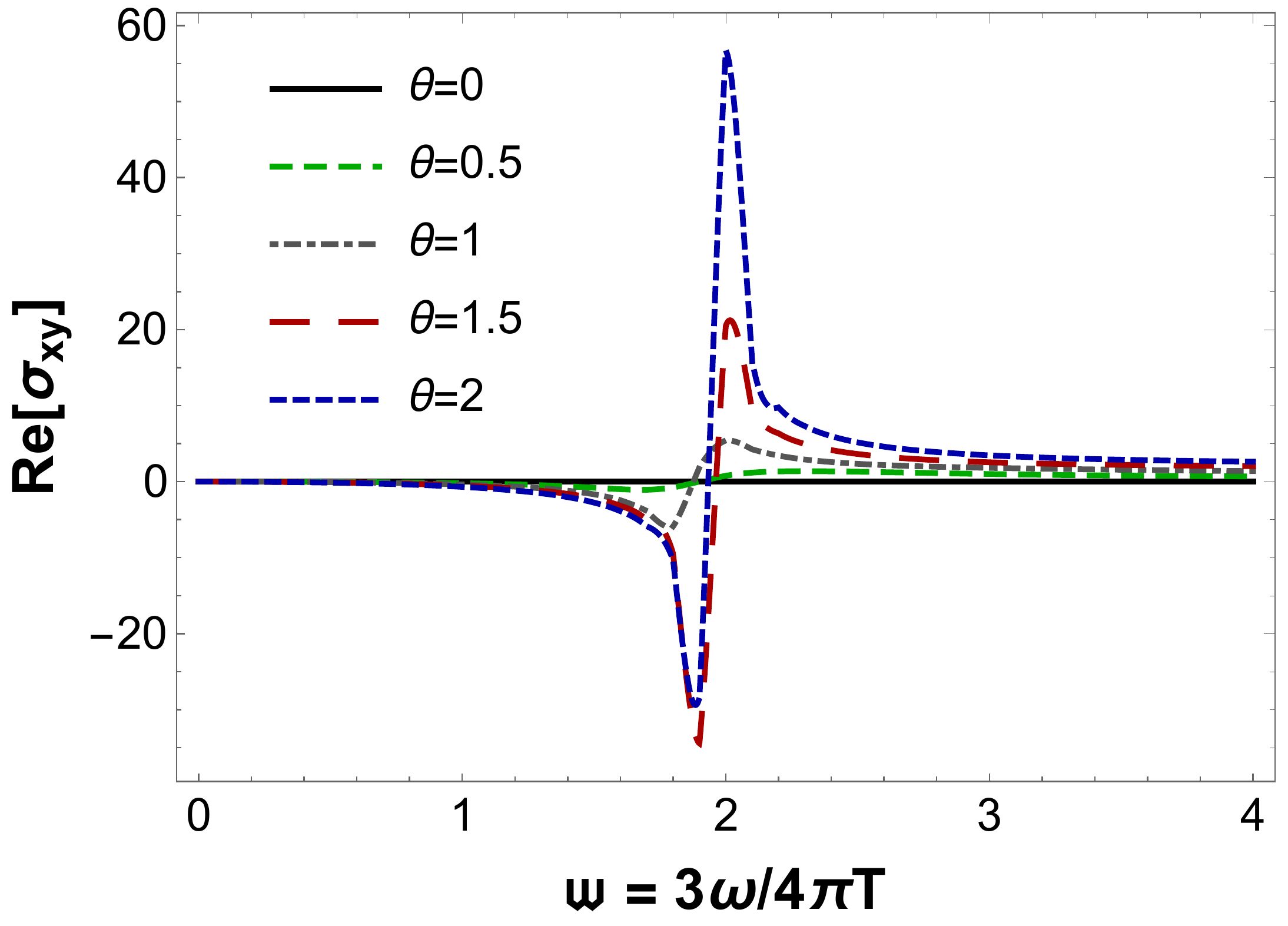}
\end{tabular}
\begin{tabular}{cc}
\includegraphics[width=0.48\textwidth]{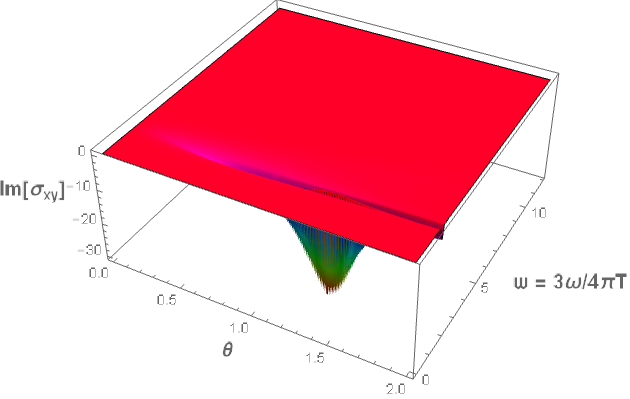} & \includegraphics[width=0.48\textwidth]{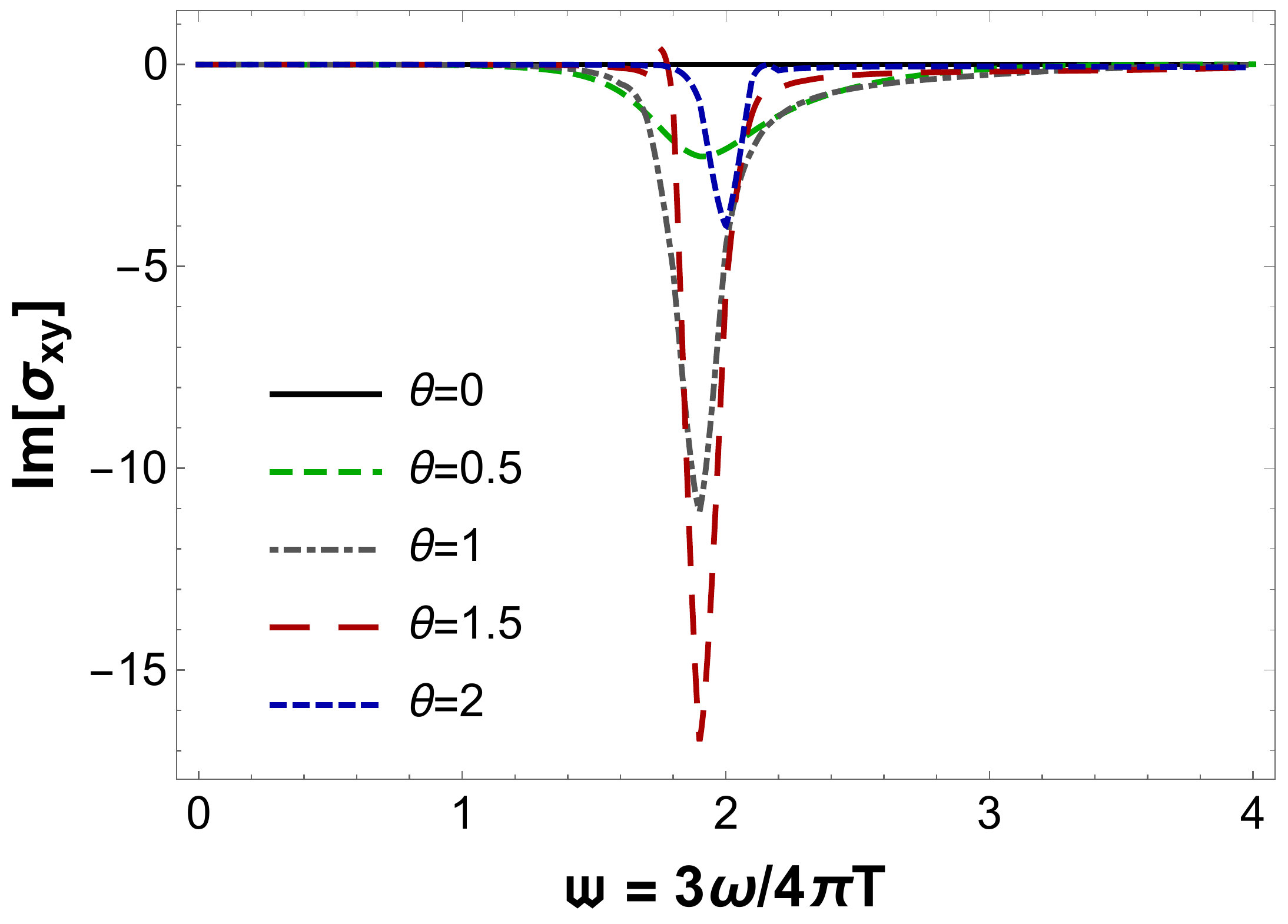}
\end{tabular}
\caption{(Color online) Real and imaginary parts of the Hall AC conductivity as functions the $\theta$-angle and the dimensionless frequency variable, $\mw=3\omega/4\pi T$, for $C_\Lambda=\Lambda L/2=2$ (dimensionless combination involving the characteristic mass scale $\Lambda$ of the bulk monopole condensate). These results were generated with the mass function $M(u)=\tanh(u)$. \label{fig4}}
\end{figure}

\begin{figure}
\begin{tabular}{cc}
\includegraphics[width=0.48\textwidth]{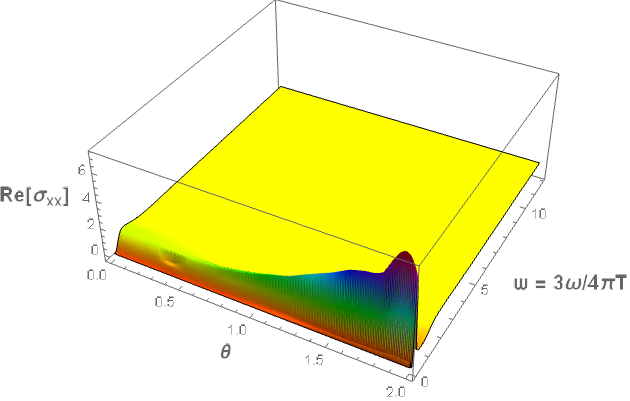} & \includegraphics[width=0.48\textwidth]{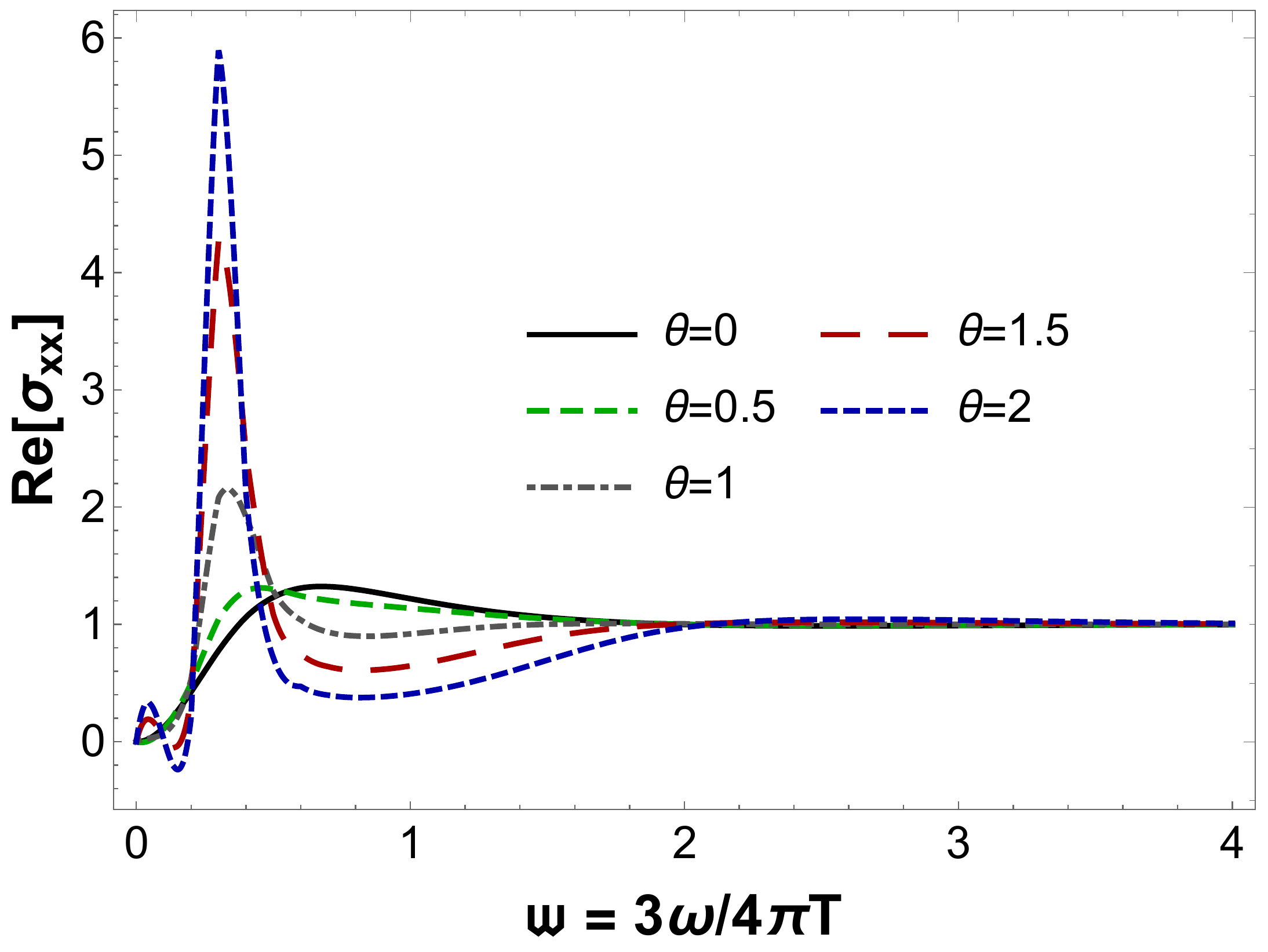}
\end{tabular}
\begin{tabular}{cc}
\includegraphics[width=0.48\textwidth]{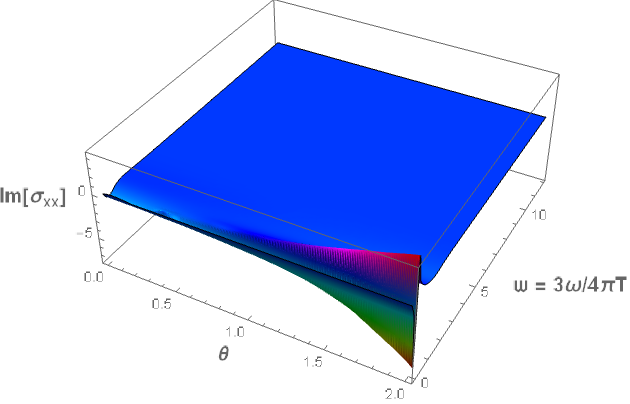} & \includegraphics[width=0.48\textwidth]{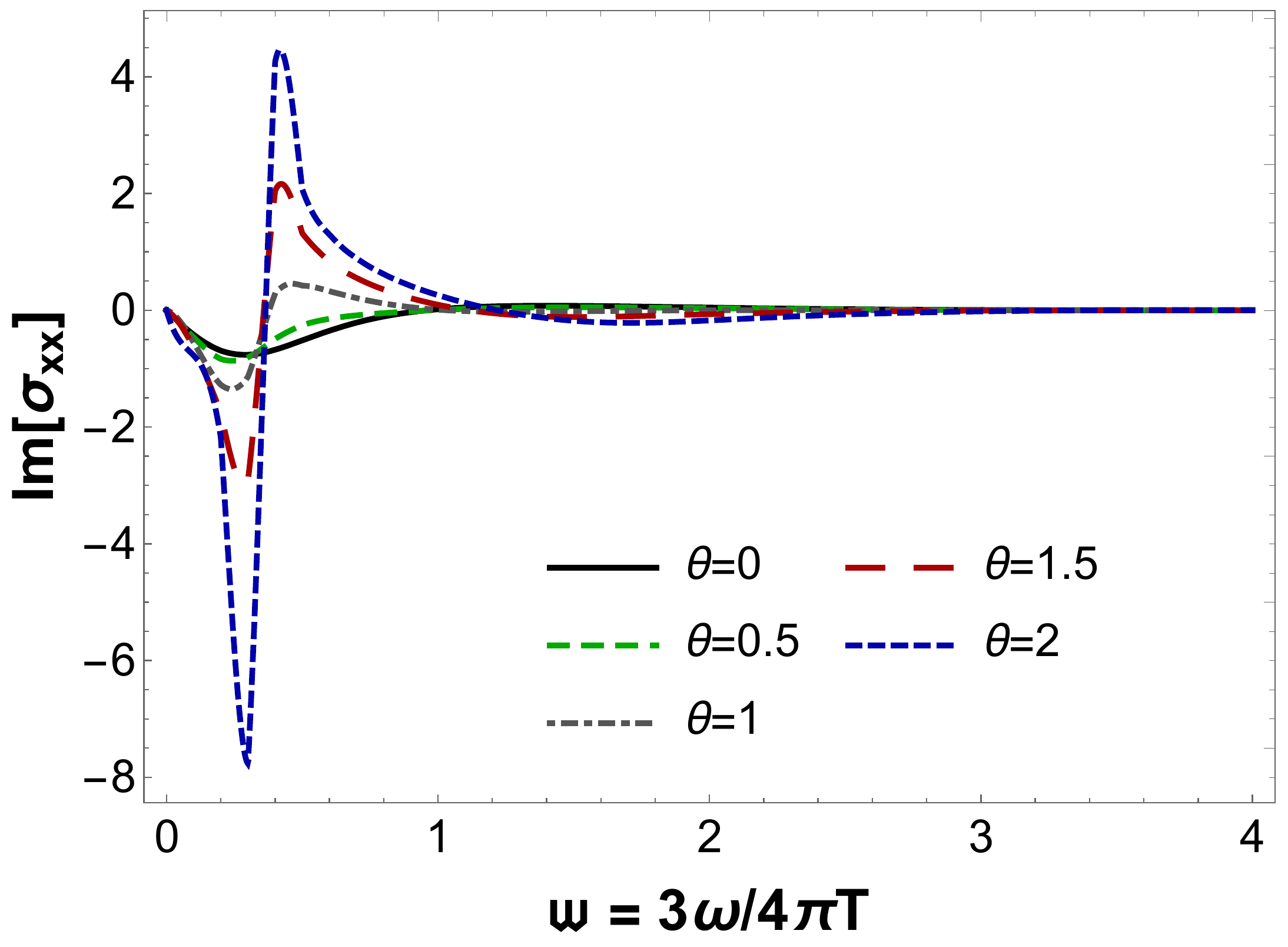}
\end{tabular}
\caption{(Color online) Real and imaginary parts of the diagonal AC conductivity as functions the $\theta$-angle and the dimensionless frequency variable, $\mw=3\omega/4\pi T$, for $C_\Lambda=\Lambda L/2=1$ (dimensionless combination involving the characteristic mass scale $\Lambda$ of the bulk monopole condensate). These results were generated with the mass function $M(u)=\tanh(u^2)$. \label{fig5}}
\end{figure}

\begin{figure}
\begin{tabular}{cc}
\includegraphics[width=0.48\textwidth]{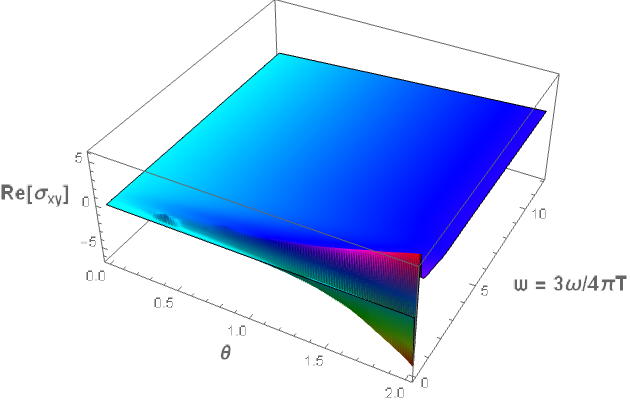} & \includegraphics[width=0.48\textwidth]{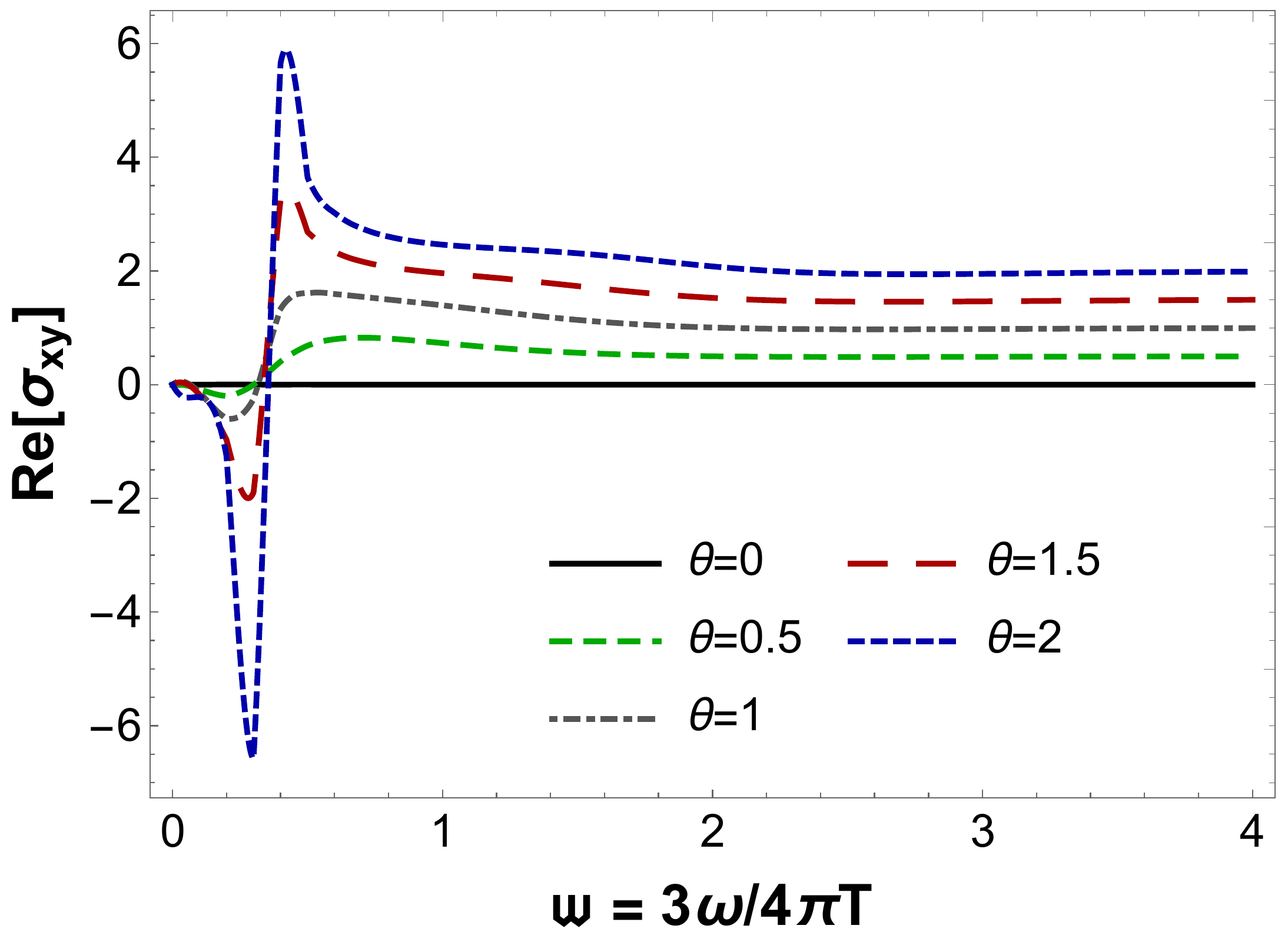}
\end{tabular}
\begin{tabular}{cc}
\includegraphics[width=0.48\textwidth]{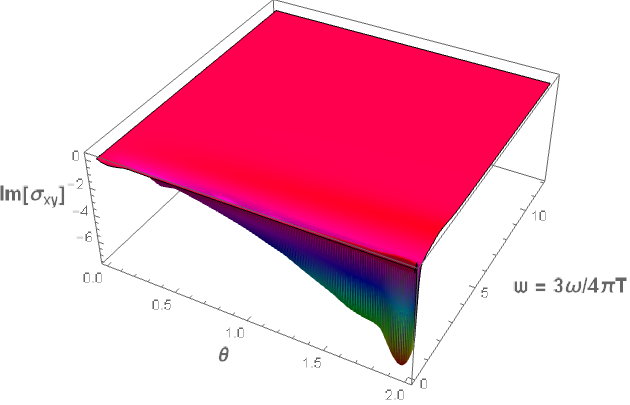} & \includegraphics[width=0.48\textwidth]{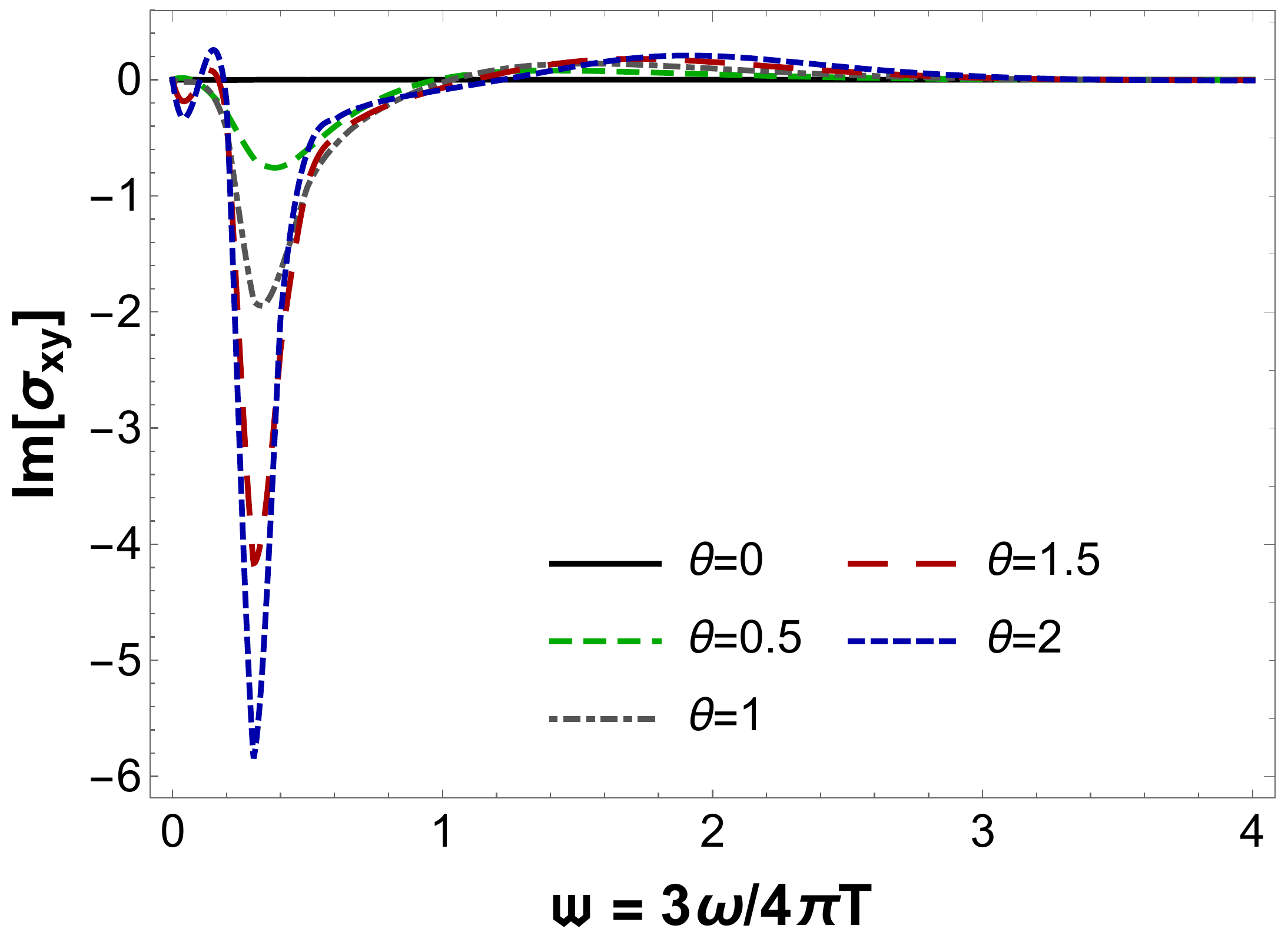}
\end{tabular}
\caption{(Color online) Real and imaginary parts of the Hall AC conductivity as functions the $\theta$-angle and the dimensionless frequency variable, $\mw=3\omega/4\pi T$, for $C_\Lambda=\Lambda L/2=1$ (dimensionless combination involving the characteristic mass scale $\Lambda$ of the bulk monopole condensate). These results were generated with the mass function $M(u)=\tanh(u^2)$. \label{fig6}}
\end{figure}

\begin{figure}
\begin{tabular}{cc}
\includegraphics[width=0.48\textwidth]{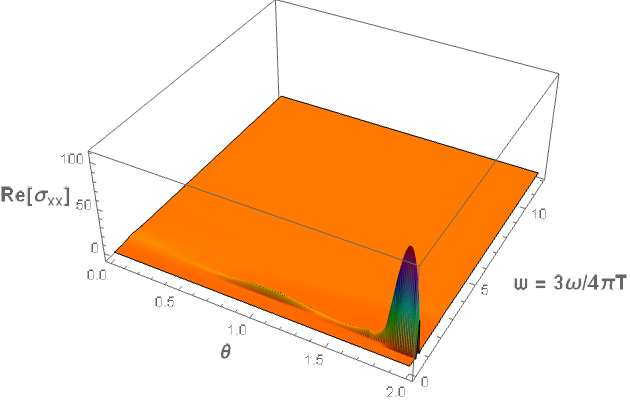} & \includegraphics[width=0.48\textwidth]{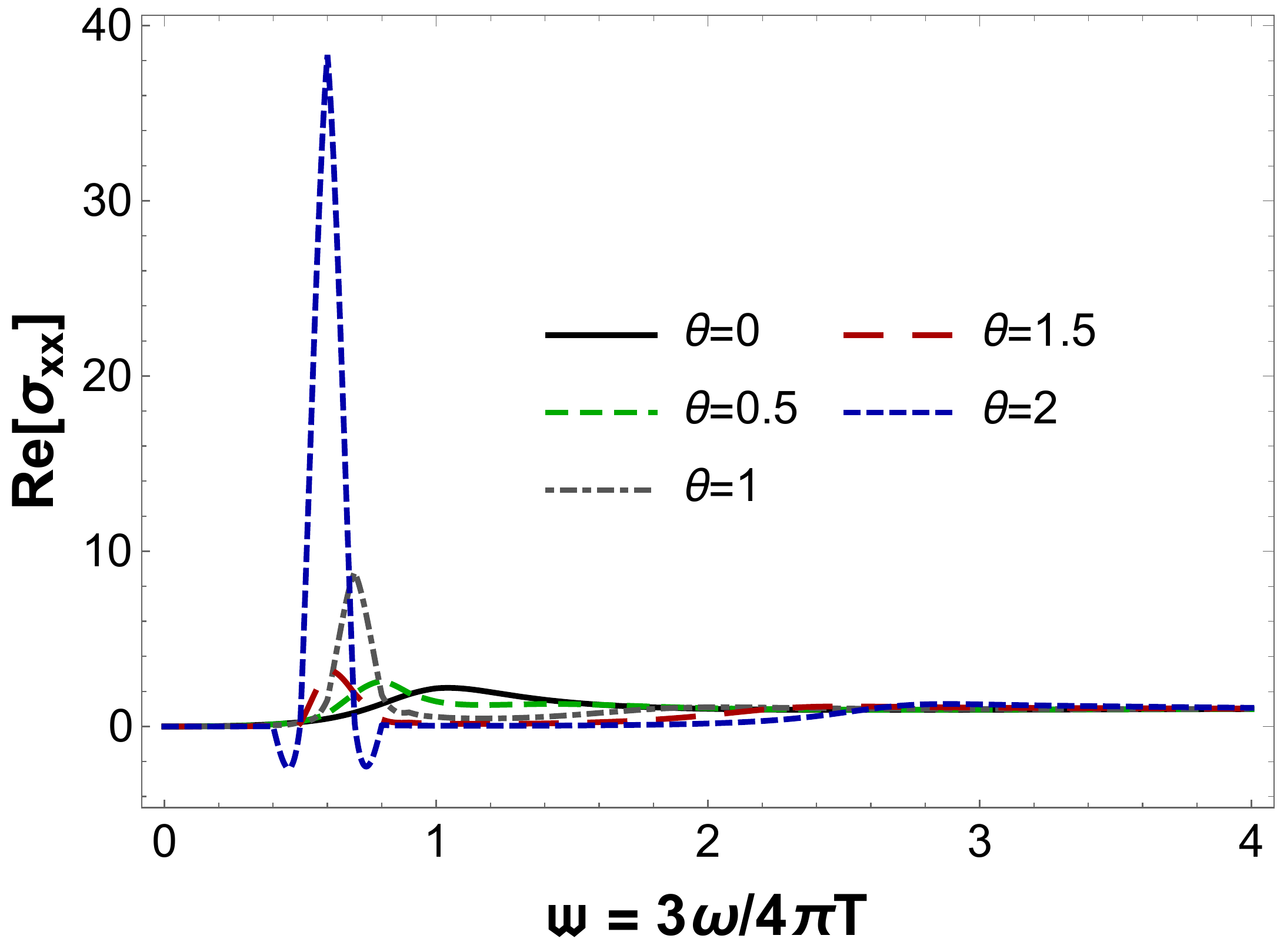}
\end{tabular}
\begin{tabular}{cc}
\includegraphics[width=0.48\textwidth]{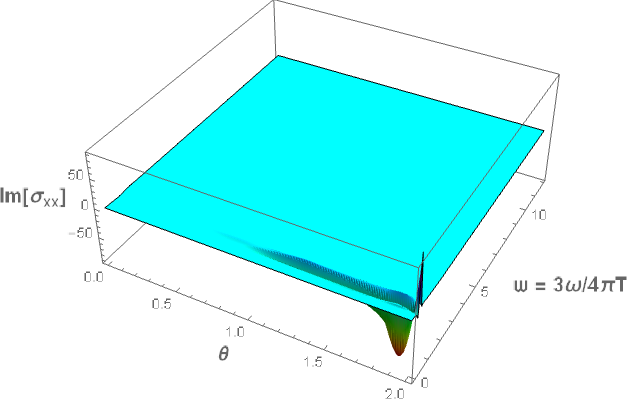} & \includegraphics[width=0.48\textwidth]{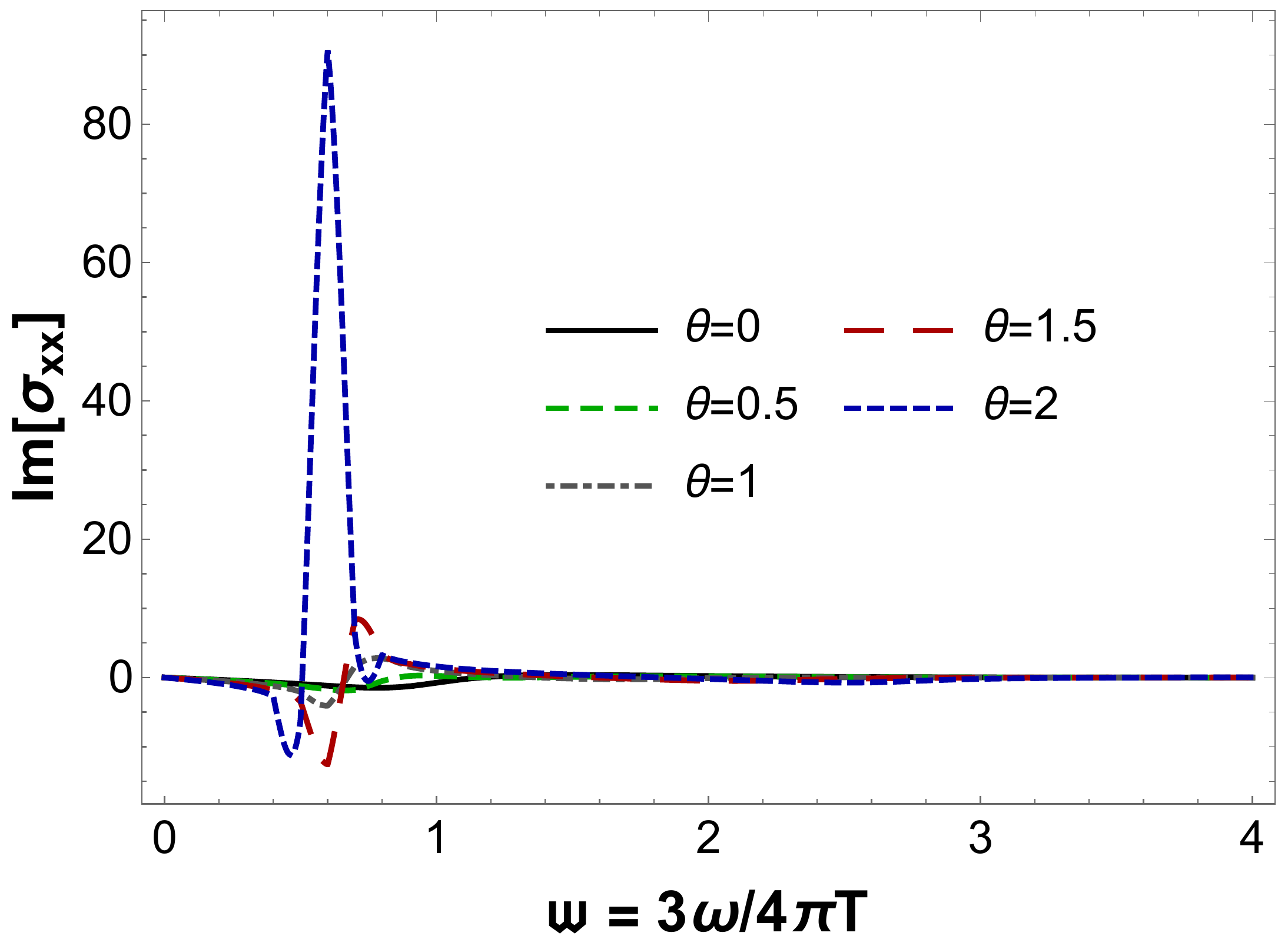}
\end{tabular}
\caption{(Color online) Real and imaginary parts of the diagonal AC conductivity as functions the $\theta$-angle and the dimensionless frequency variable, $\mw=3\omega/4\pi T$, for $C_\Lambda=\Lambda L/2=2$ (dimensionless combination involving the characteristic mass scale $\Lambda$ of the bulk monopole condensate). These results were generated with the mass function $M(u)=\tanh(u^2)$. \label{fig7}}
\end{figure}

\begin{figure}
\begin{tabular}{cc}
\includegraphics[width=0.48\textwidth]{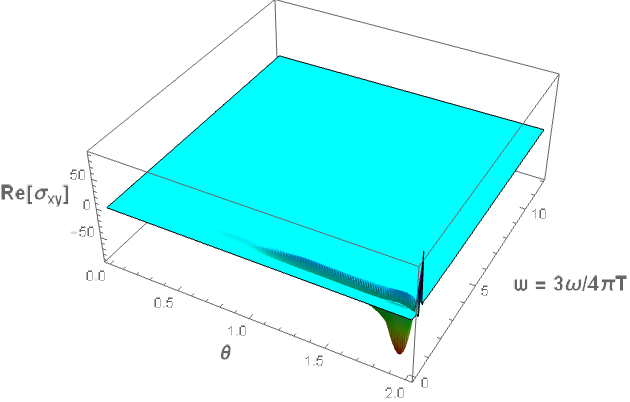} & \includegraphics[width=0.48\textwidth]{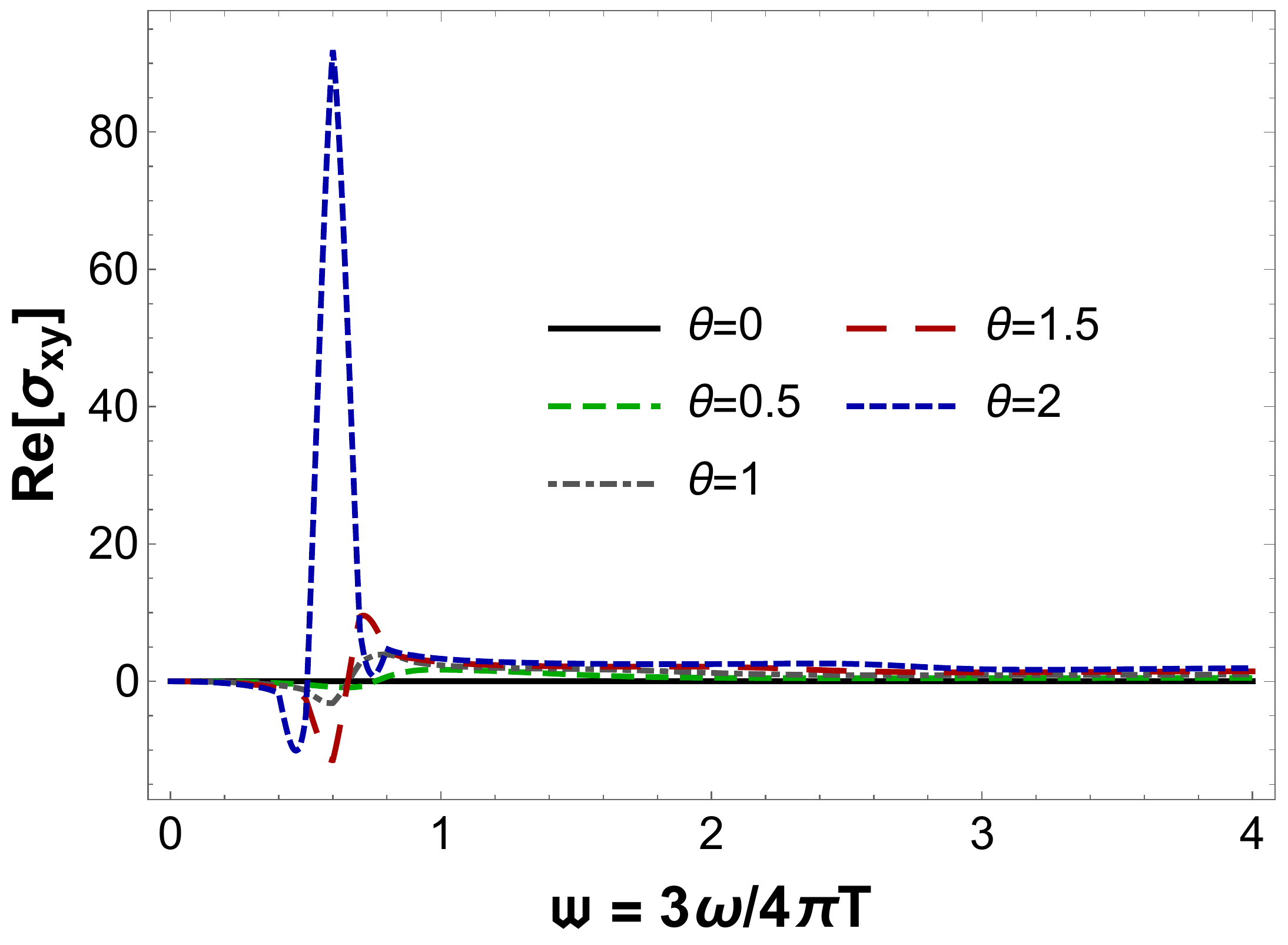}
\end{tabular}
\begin{tabular}{cc}
\includegraphics[width=0.48\textwidth]{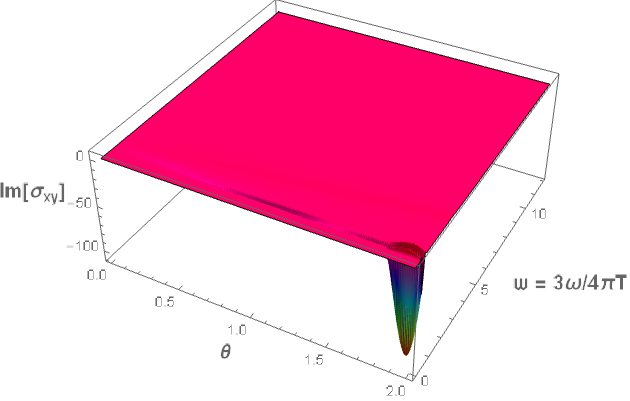} & \includegraphics[width=0.48\textwidth]{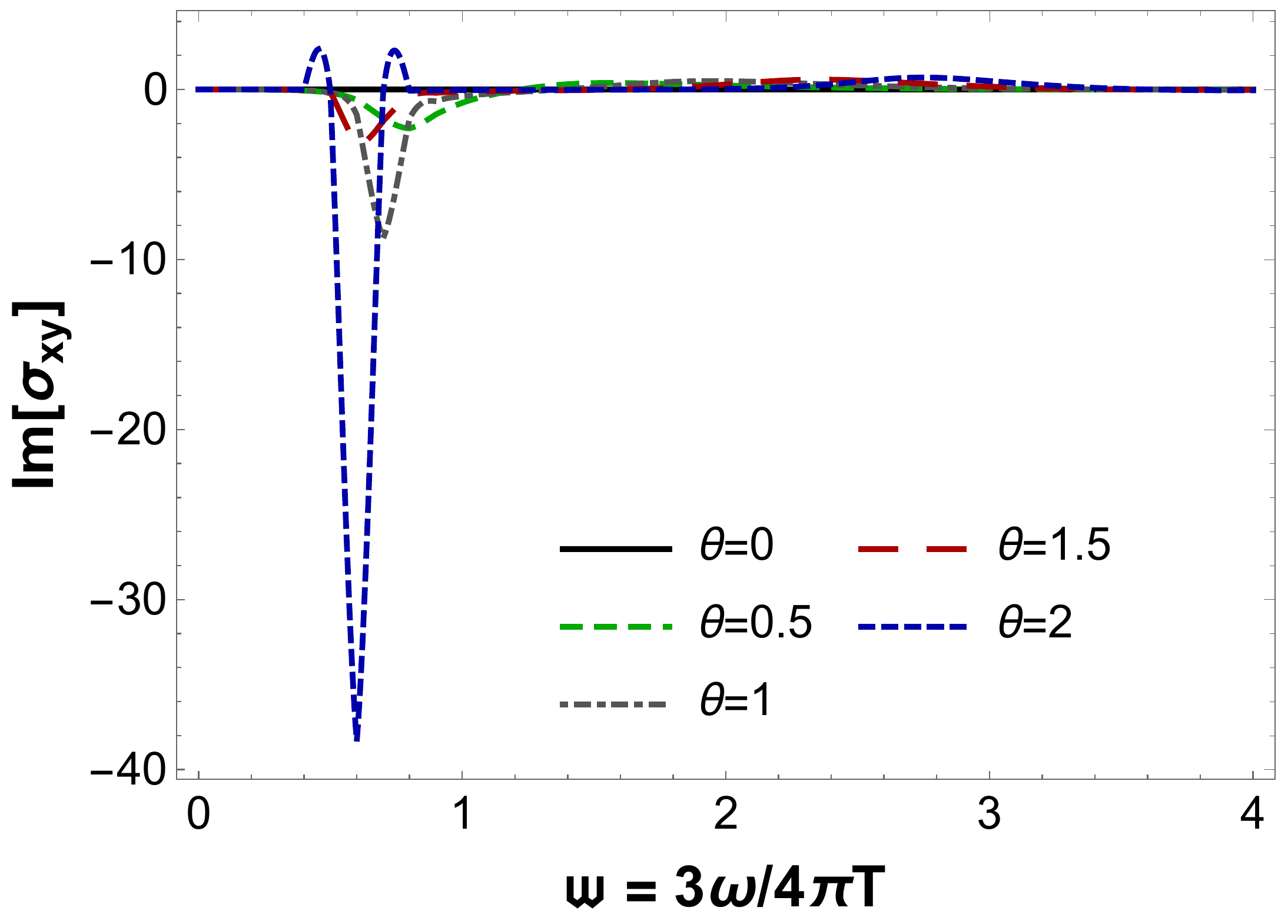}
\end{tabular}
\caption{(Color online) Real and imaginary parts of the Hall AC conductivity as functions the $\theta$-angle and the dimensionless frequency variable, $\mw=3\omega/4\pi T$, for $C_\Lambda=\Lambda L/2=2$ (dimensionless combination involving the characteristic mass scale $\Lambda$ of the bulk monopole condensate). These results were generated with the mass function $M(u)=\tanh(u^2)$. \label{fig8}}
\end{figure}

\begin{figure}
\begin{tabular}{cc}
\includegraphics[width=0.48\textwidth]{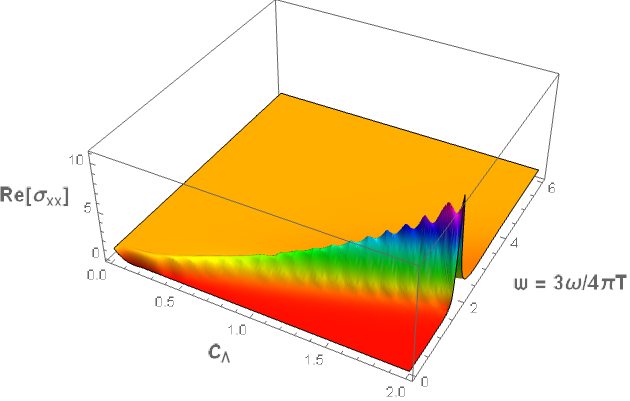} & \includegraphics[width=0.48\textwidth]{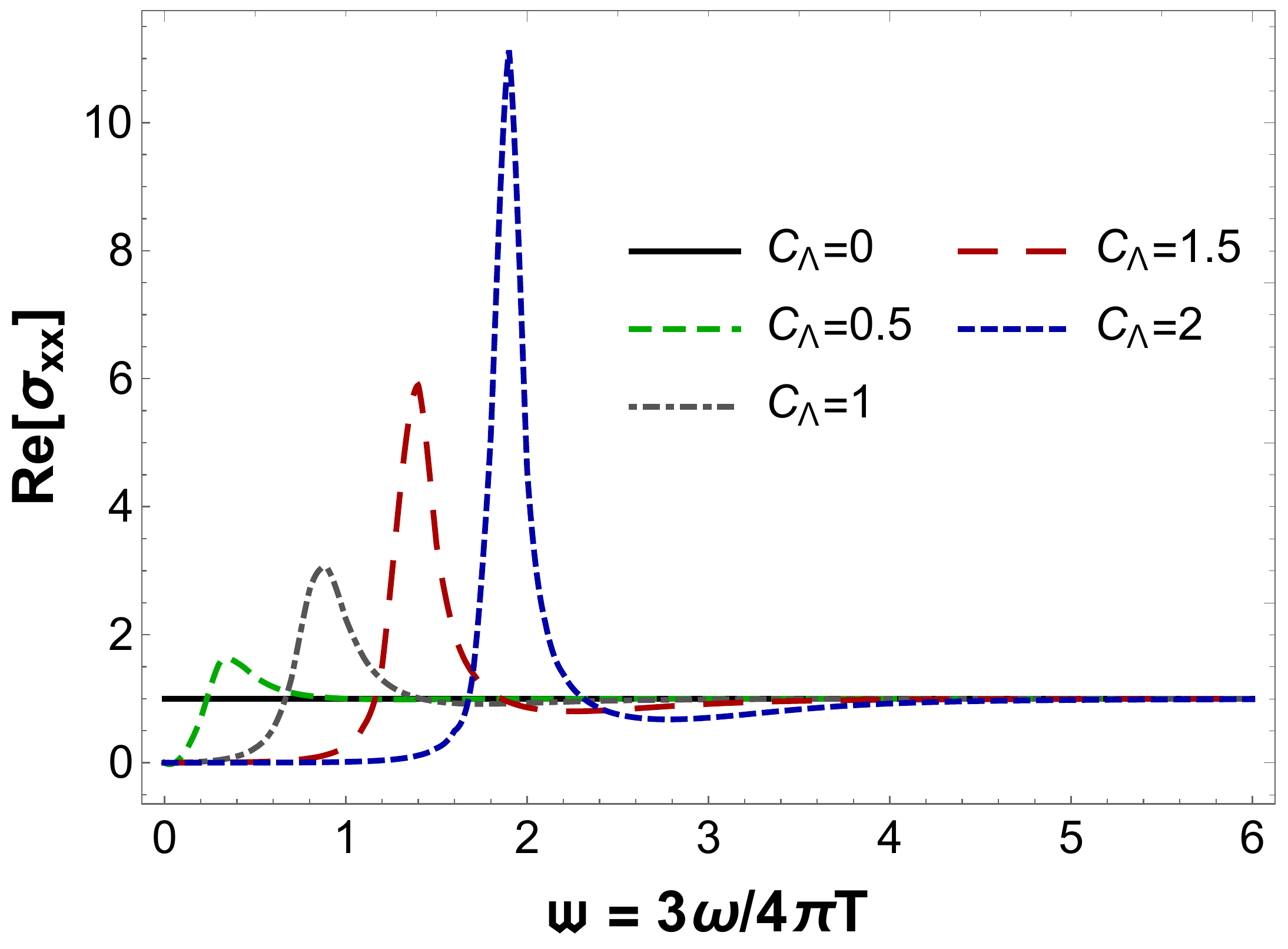}
\end{tabular}
\begin{tabular}{cc}
\includegraphics[width=0.48\textwidth]{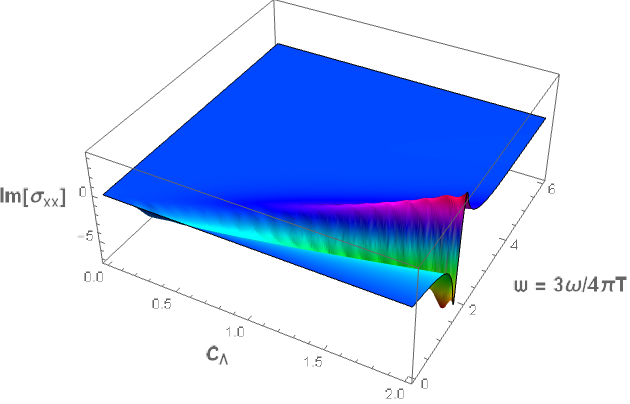} & \includegraphics[width=0.48\textwidth]{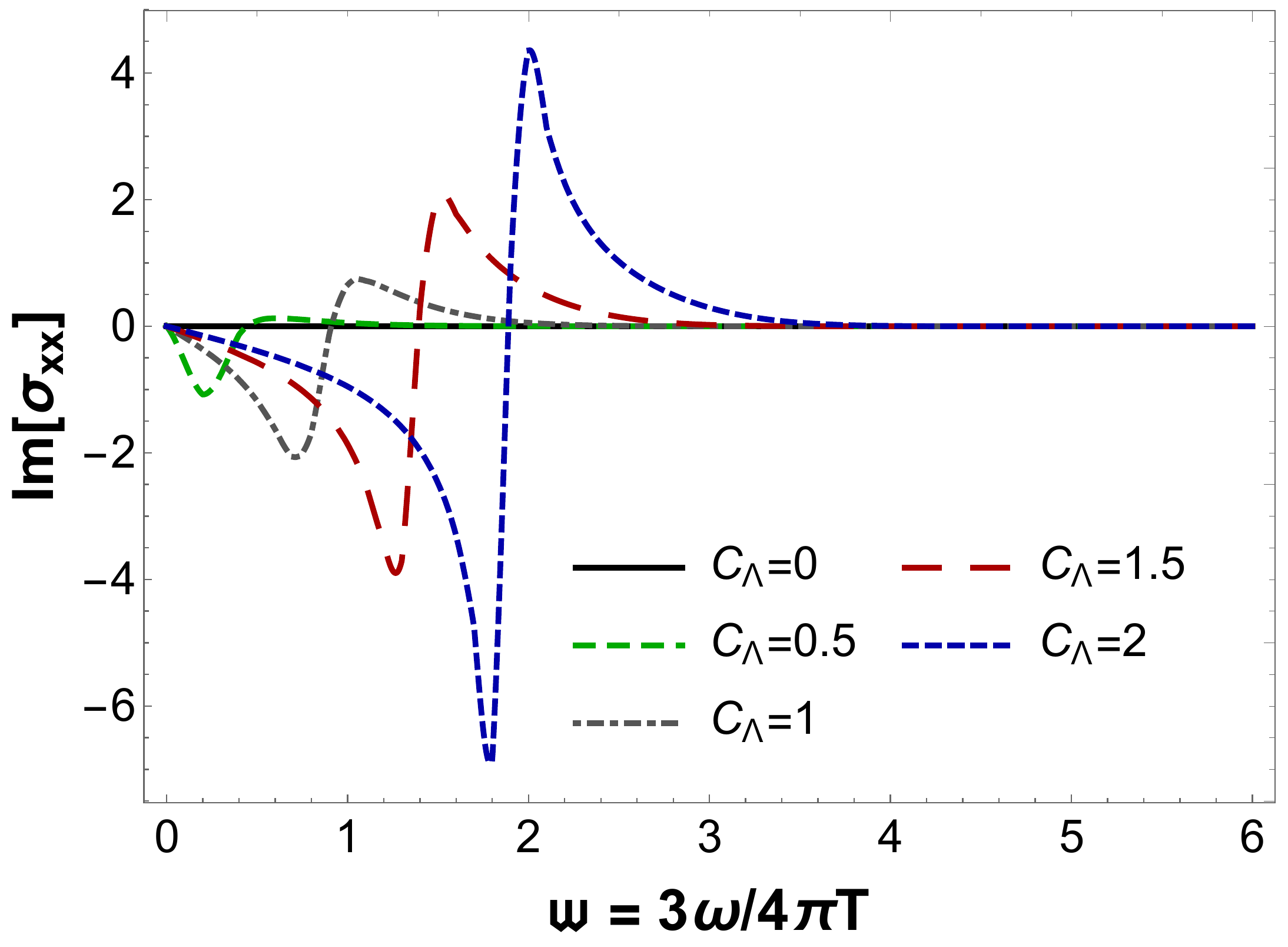}
\end{tabular}
\caption{(Color online) Real and imaginary parts of the diagonal AC conductivity as functions of $(C_\Lambda=\Lambda L/2,\mw=3\omega/4\pi T)$ at fixed $\theta=1$. These results were generated with the mass function $M(u)=\tanh(u)$. \label{fig9}}
\end{figure}

\begin{figure}
\begin{tabular}{cc}
\includegraphics[width=0.48\textwidth]{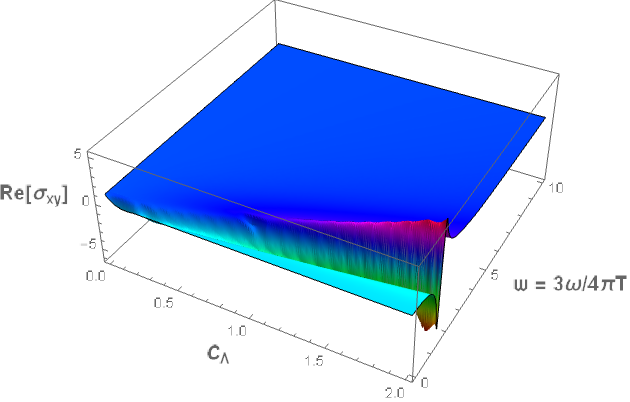} & \includegraphics[width=0.48\textwidth]{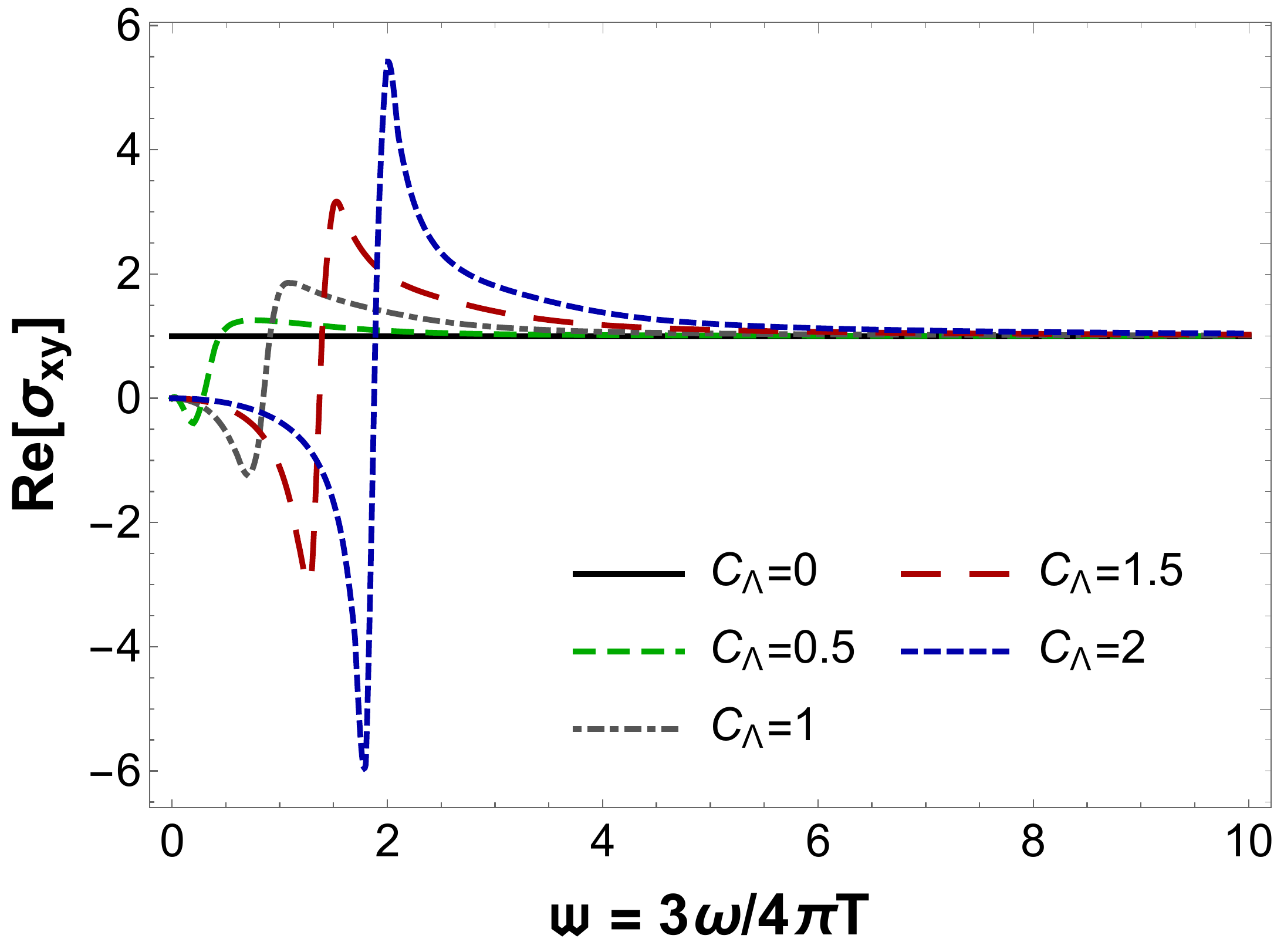}
\end{tabular}
\begin{tabular}{cc}
\includegraphics[width=0.48\textwidth]{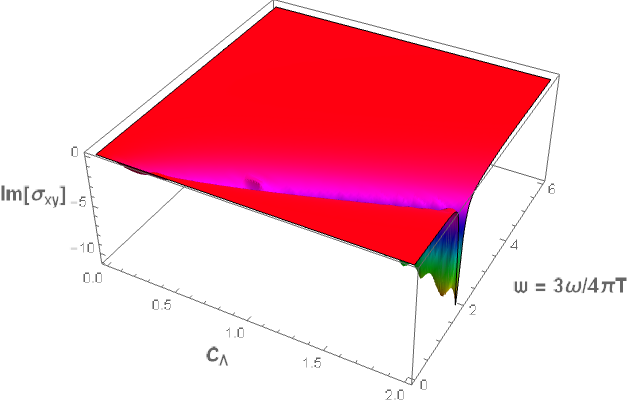} & \includegraphics[width=0.48\textwidth]{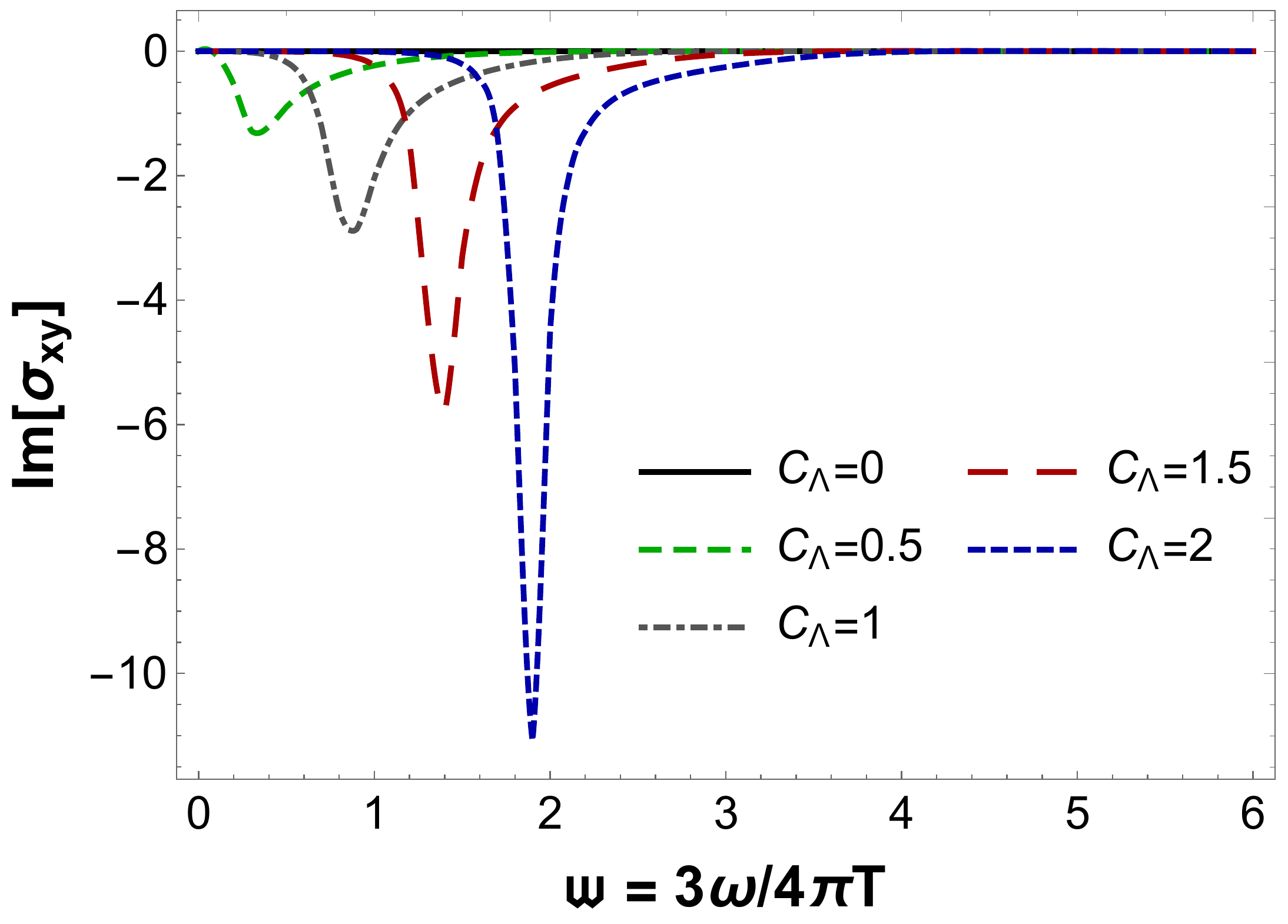}
\end{tabular}
\caption{(Color online) Real and imaginary parts of the Hall AC conductivity as functions of $(C_\Lambda=\Lambda L/2,\mw=3\omega/4\pi T)$ at fixed $\theta=1$. These results were generated with the mass function $M(u)=\tanh(u)$. \label{fig10}}
\end{figure}

\begin{figure}
\begin{tabular}{cc}
\includegraphics[width=0.48\textwidth]{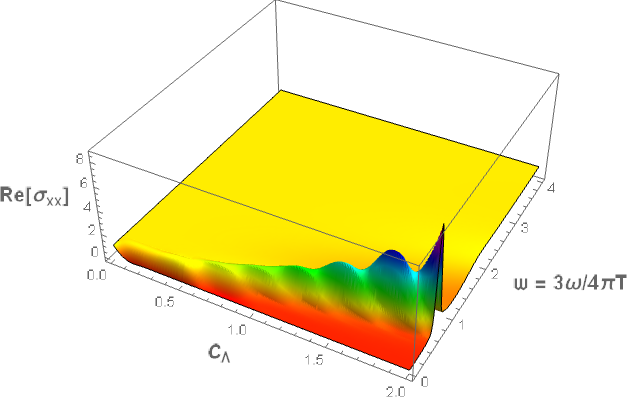} & \includegraphics[width=0.48\textwidth]{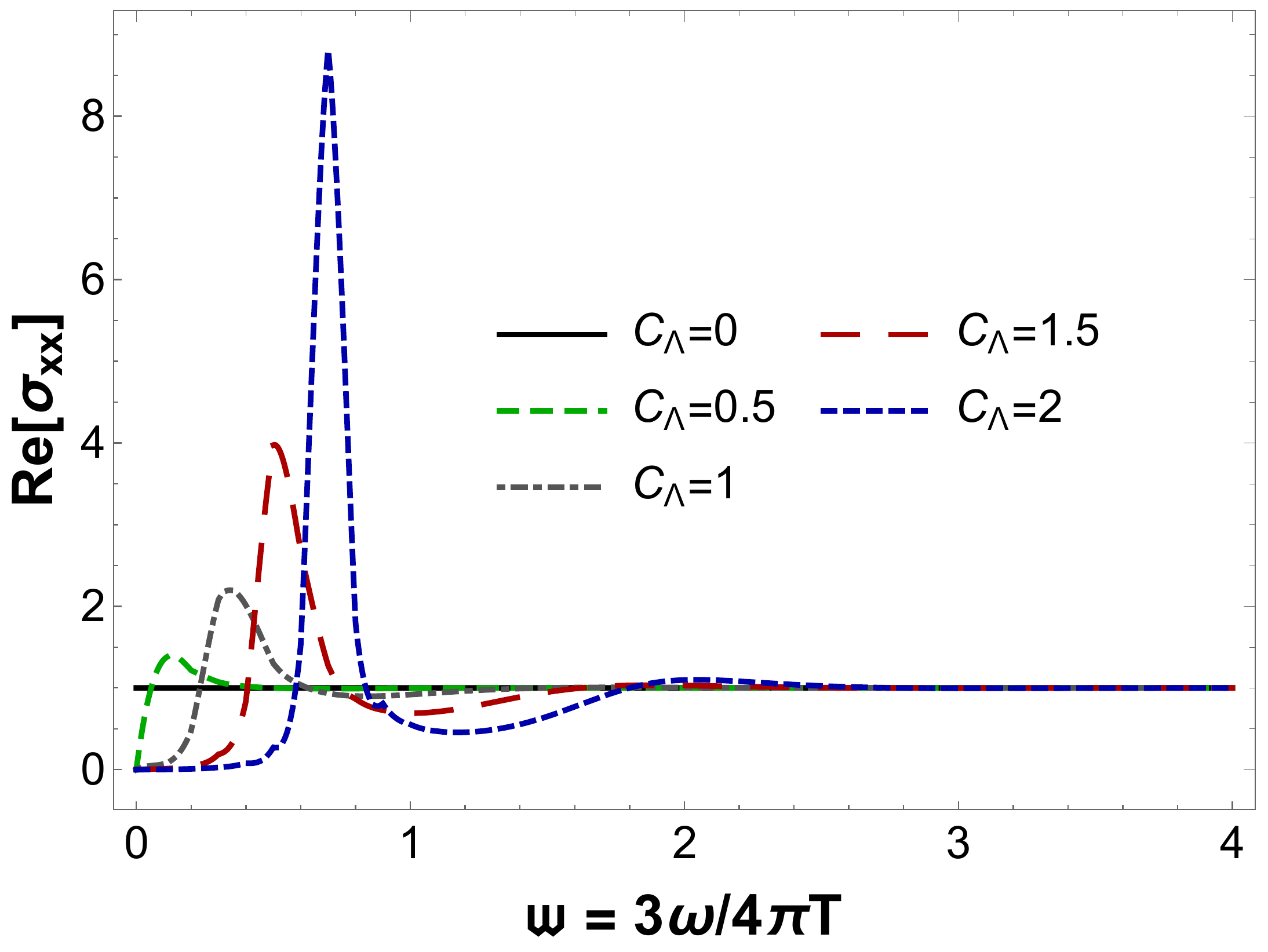}
\end{tabular}
\begin{tabular}{cc}
\includegraphics[width=0.48\textwidth]{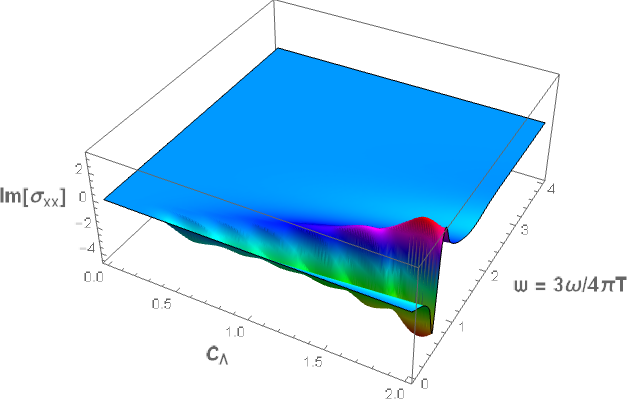} & \includegraphics[width=0.48\textwidth]{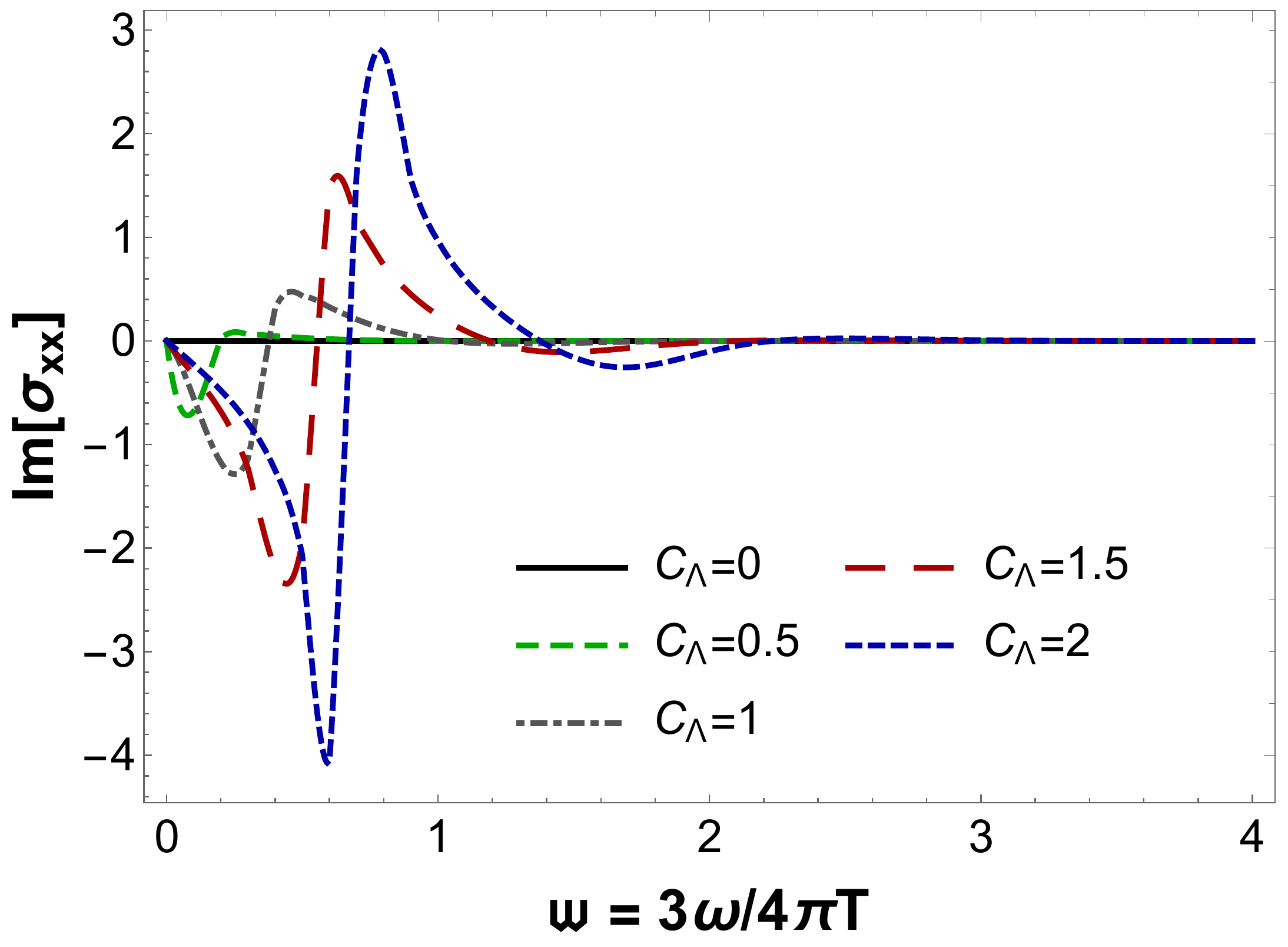}
\end{tabular}
\caption{(Color online) Real and imaginary parts of the diagonal AC conductivity as functions of $(C_\Lambda=\Lambda L/2,\mw=3\omega/4\pi T)$ at fixed $\theta=1$. These results were generated with the mass function $M(u)=\tanh(u^2)$. \label{fig11}}
\end{figure}

\begin{figure}
\begin{tabular}{cc}
\includegraphics[width=0.48\textwidth]{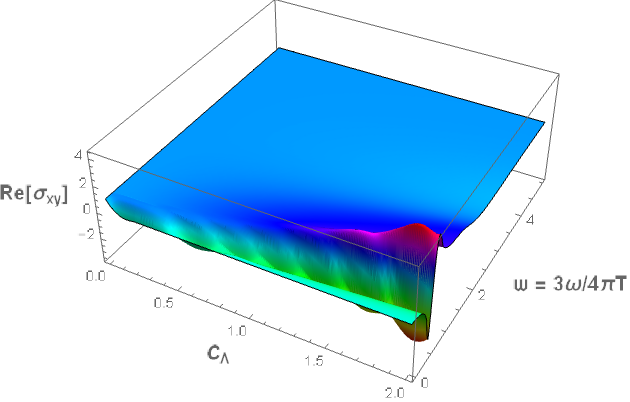} & \includegraphics[width=0.48\textwidth]{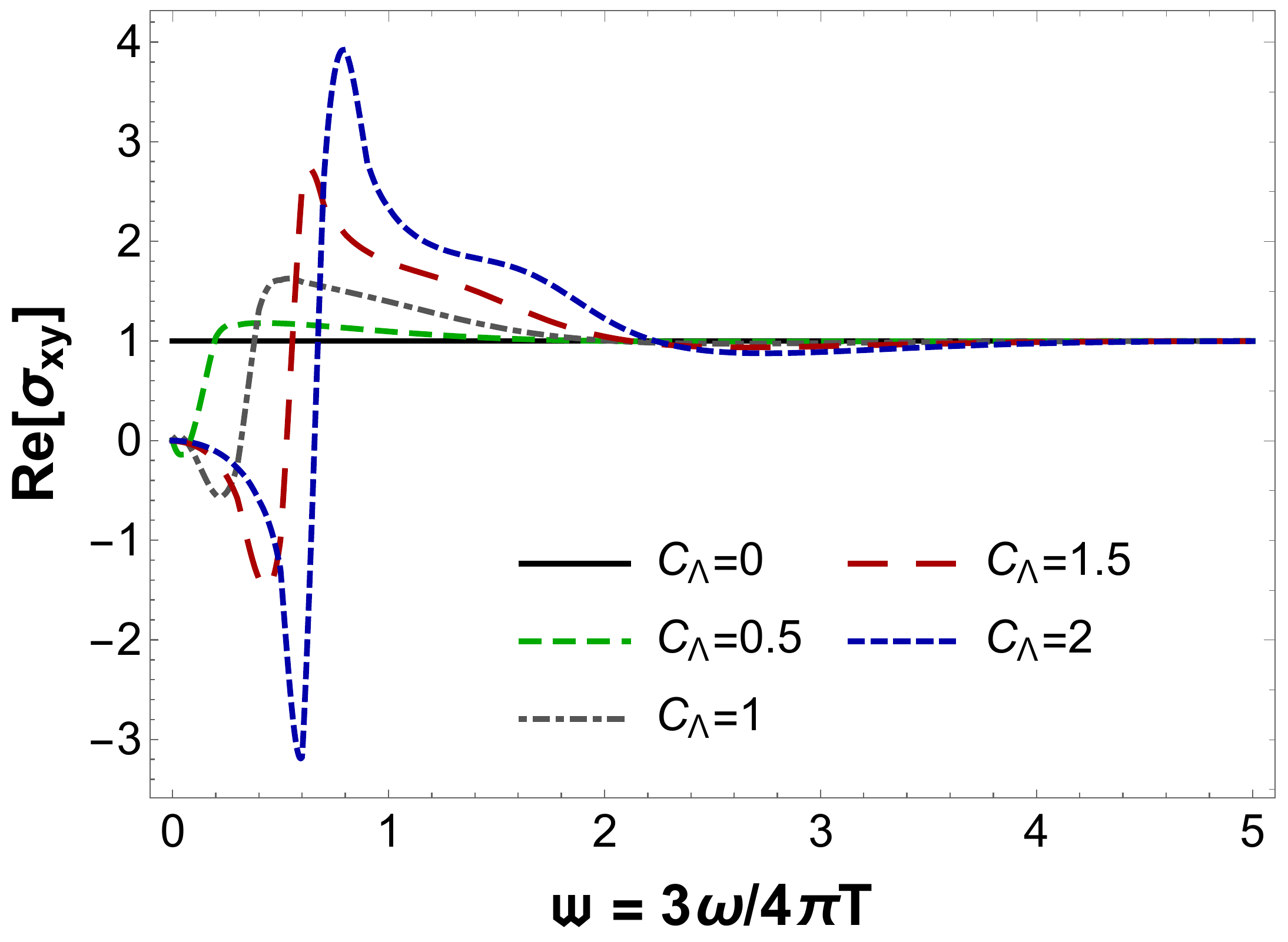}
\end{tabular}
\begin{tabular}{cc}
\includegraphics[width=0.48\textwidth]{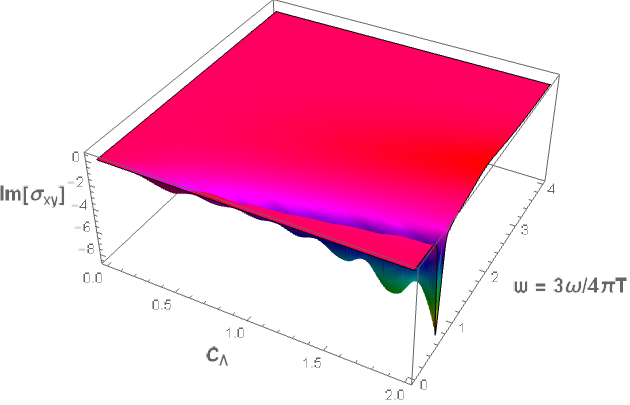} & \includegraphics[width=0.48\textwidth]{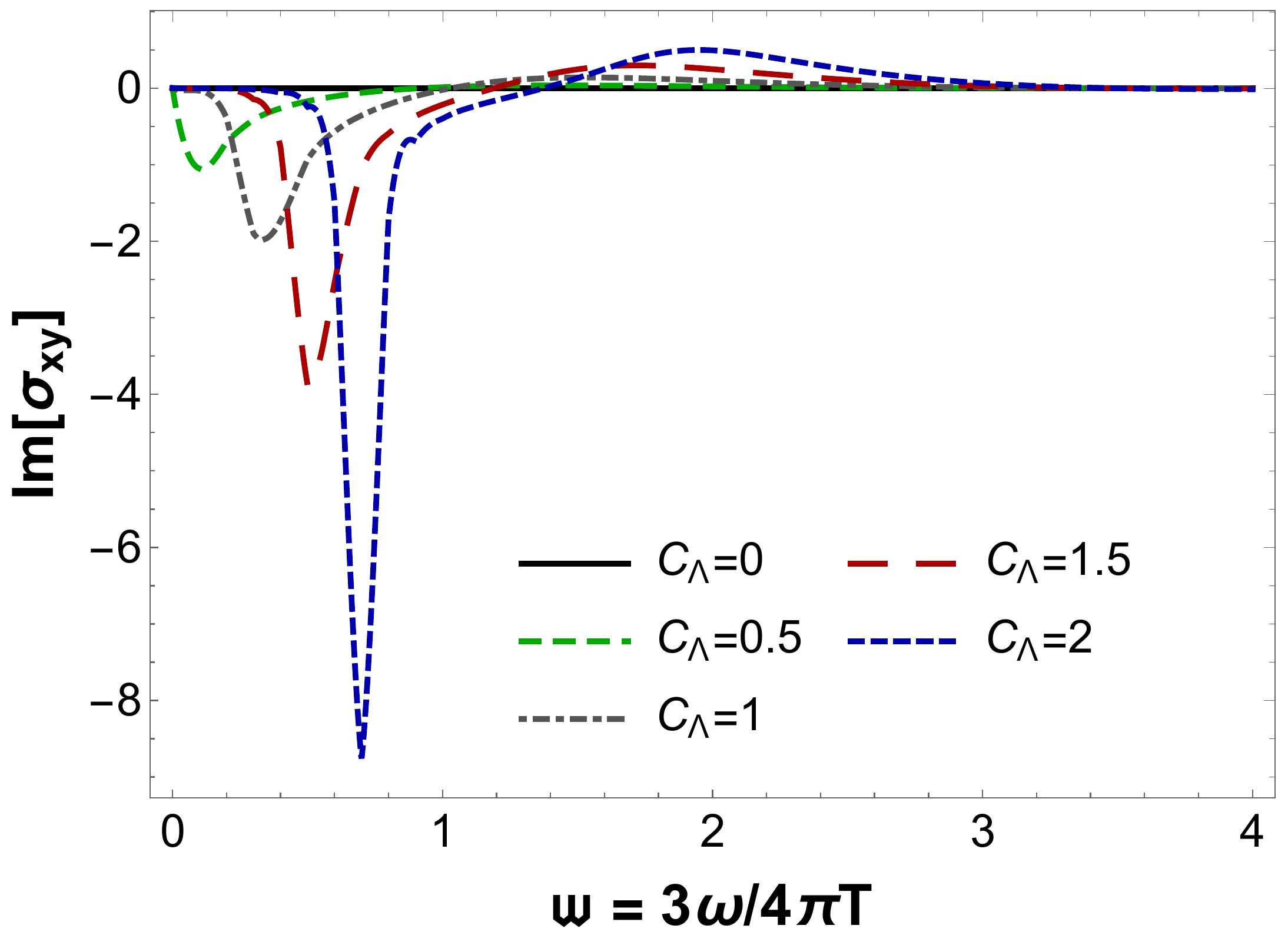}
\end{tabular}
\caption{(Color online) Real and imaginary parts of the Hall AC conductivity as functions of $(C_\Lambda=\Lambda L/2,\mw=3\omega/4\pi T)$ at fixed $\theta=1$. These results were generated with the mass function $M(u)=\tanh(u^2)$. \label{fig12}}
\end{figure}

\subsubsection{Holographic operator mixing}
\label{sec:mix}

The ultraviolet asymptotics of the massive Kalb-Ramond field subjected to the boundary condition \eqref{KRbc} is the very same one of the Maxwell field,\footnote{As long as the prescribed profile for the effective mass field $m(u)$ is chosen such that $m(u\to\epsilon)\sim\epsilon^a$, with $a=1$ or $a>3/2$, as originally discussed in Ref. \cite{Rougemont:2015gia}.} which in a four dimensional bulk goes like,
\begin{align}
K_{tx(y)}(u\to 0,\omega) = S(\omega)+O(\omega)u+\cdots,
\label{eqasymp}
\end{align}
therefore, it already goes to a constant at the boundary and no field redefinitions are needed on the lines of the method discussed in Ref. \cite{Kim:2014bza}. The effective action \eqref{SKR5} may be recast in the following form,
\begin{align}
\mathcal{S}_{\textrm{cond}}^{\theta,\textrm{bdy}}[K_{tx}^0,K_{ty}^0]&=-\frac{1}{2}\int\frac{d\omega}{2\pi}\left\{- \lim_{u=\epsilon\to 0} \left[K_{a}^*\mathbb{A}_{ab}K_b+K_{a}^*\mathbb{B}_{ab}K'_{b} \right]\biggr|_{\textrm{on-shell}}^{\textrm{infalling}}\right\},
\label{SKR6}
\end{align}
where the indexes $a,b\in\left\{1\equiv tx,2\equiv ty\right\}$ and,
\begin{align}
\mathbb{A}_{ab}(u,\omega)\equiv\frac{i\omega\theta\,\varepsilon_{ab}}{\omega^2-m^2(u)g_{tt}},\qquad
\mathbb{B}_{ab}(u,\omega)\equiv\sqrt{\frac{g_{tt}}{g_{uu}}}\frac{\delta_{ab}}{\omega^2-m^2(u)g_{tt}}.
\end{align}

Now one needs to expand the fields $K_a=\left(K_{tx},K_{ty}\right)^T$ near the horizon at $u=u_H$ and impose the infalling wave condition, whose ansatz reads as follows,
\begin{align}
K_a(u\to u_H,\omega) = (u_H-u)^{-i\omega/4\pi T}\left[\varphi_a(\omega)+\tilde{\varphi}_a(\omega)(u_H-u)+\cdots\right],
\label{IRexp}
\end{align}
where with the choice of the infalling wave condition at the horizon I fixed two boundary conditions, and there are still two boundary conditions to be fixed at the horizon (since there are two second order coupled ODE's for $K_{tx}$ and $K_{ty}$).\footnote{Notice that by taking the infalling wave condition one sets to zero the two leading coefficients close to the horizon corresponding to the outgoing modes. The other two leading coefficients that remain to be fixed are the $\varphi_a(\omega)$ in \eqref{IRexp}. The subleading coefficients $\tilde{\varphi}_a(\omega)$ (and also the other subsubleading coefficients omitted in the expansion) can be fixed in terms of $\varphi_a(\omega)$ by substituting the infrared expansion \eqref{IRexp} back into the equations of motion \eqref{Ktxeom} and \eqref{Ktyeom}, and then setting to zero each power of $(u_H-u)$ in the resulting algebraic equations.} This is done by fixing two linearly independent sets of initial conditions $\varphi_{a(i)}$, $(i)=1,2$, which can be chosen as follows \cite{Kim:2014bza},\footnote{As discussed in Ref. \cite{Kim:2014bza}, the results are independent of the specific chosen sets, the only requirement is that these sets must be linearly independent.}
\begin{align}
\varphi_{a(1)}=(1,1)^T=
\begin{pmatrix}
1\\
1
\end{pmatrix},\qquad \varphi_{a(2)}=(1,-1)^T=
\begin{pmatrix}
1\\
-1
\end{pmatrix},
\label{ICs}
\end{align}
where the index $a\in\left\{1\equiv tx,2\equiv ty\right\}$ denotes the line and the index $(i)=1,2$ denotes the column of the $2\times 2$ matrix of initial conditions,
\begin{align}
\varphi_{a(i)}=\left(\varphi_{a(1)},\varphi_{a(2)}\right)=
\begin{pmatrix}
1 & 1\\
1 & -1
\end{pmatrix}.
\label{ICs-2}
\end{align}

Near the boundary ($u=\epsilon\to 0$), according to Eq. \eqref{eqasymp}, the solutions are then expanded as follows,
\begin{align}
K_{a(i)}(u\to 0,\omega) = \mathbb{S}_{a(i)}(\omega)+\mathbb{O}_{a(i)}(\omega)u+\cdots,
\label{UVsols}
\end{align}
and the general solution is then a linear combination of the solutions \eqref{UVsols},
\begin{align}
K_a(u,\omega) = K_{a(i)}(u,\omega)c_{(i)}(\omega).
\label{UVsols-2}
\end{align}
The (radial) constants $c_{(i)}$ must be chosen such that the combined sources coincide with the boundary values $J_a$ of the bulk fields,
\begin{align}
J_a=\mathbb{S}_{a(i)}c_{(i)} \Rightarrow
\begin{pmatrix}
K_{tx}^0\\
K_{ty}^0
\end{pmatrix}=
\begin{pmatrix}
\mathbb{S}_{1(1)} & \mathbb{S}_{1(2)}\\
\mathbb{S}_{2(1)} & \mathbb{S}_{2(2)}
\end{pmatrix}
\begin{pmatrix}
c_{(1)}\\
c_{(2)}
\end{pmatrix}.
\label{eqbdysc}
\end{align}

Thus, one needs to read off the answer of $K'_a(u\to 0)\sim\mathbb{O}_{a(i)}c_{(i)}$ with respect to $J_a$, that is, one needs to figure out how the derivatives $K'_{tx(y)}$ relate to the sources $K_{tx(y)}^0$. When there are no constraints regarding some residual diffeomorphism invariance (as in the present case), one can fix $c_{(i)}$ by simply inverting Eq. \eqref{eqbdysc}, since in this case $\textrm{det}(\mathbb{S})\neq 0$ \cite{Kim:2014bza}, therefore,
\begin{align}
c_{(i)} = \left(\mathbb{S}^{-1}\right)_{(i)a}J_a.
\label{eqinv}
\end{align}
In this way, one may write down the following relation close to the boundary,
\begin{align}
\mathbb{B}_{ab}K'_b(u\to 0) = \mathbb{B}_{ab}\mathbb{O}_{b(i)}\left(\mathbb{S}^{-1}\right)_{(i)d}J_d,
\label{eqmix}
\end{align}
where, from Eq. \eqref{UVsols}, $\mathbb{S}_{a(i)}=K_{a(i)}(u=0)$ and $\mathbb{O}_{a(i)}=K'_{a(i)}(u=0)$. Notice the numerical integration of the coupled equations of motion \eqref{Ktxeom} and \eqref{Ktyeom} must be done from the horizon to the boundary. If no relevant constraints are being inadvertently neglected, when substituting the infrared expansions \eqref{IRexp} into the equations of motion and then fixing the expansion coefficients as discussed before, there should be left unspecified two independent coefficients in this infrared analysis \cite{Kim:2014bza}, $\varphi_a=\varphi_1=\varphi_{tx}$ and $\varphi_a=\varphi_2=\varphi_{ty}$. These coefficients are then fixed by choosing two different sets of initial conditions according to Eqs. \eqref{ICs}, which will then generate the required numerical solutions $K_{a(i)}$.

From the above developments, one can now rewrite the effective action \eqref{SKR6} as follows,
\begin{align}
\mathcal{S}_{\textrm{cond}}^{\theta,\textrm{bdy}}[K_{tx}^0,K_{ty}^0]&=-\frac{1}{2}\int\frac{d\omega}{2\pi}J_a^*\left\{- \lim_{u=\epsilon\to 0} \left[\mathbb{A}_{ad}+\mathbb{B}_{ab}\mathbb{O}_{b(i)}\left(\mathbb{S}^{-1}\right)_{(i)d} \right]\biggr|_{\textrm{on-shell}}^{\textrm{infalling}}\right\}J_d,
\label{SKR7}
\end{align}
and since $J_a=\left(K_{tx}^0,K_{ty}^0\right)^T=-i\omega\left(A_{x}^0,A_{y}^0\right)^T$, one rewrites the above result in terms of the sources $A_{x(y)}^0$ of the conserved vector currents at the boundary QFT in the magnetically condensed phase as,
\begin{align}
\mathcal{S}_{\textrm{cond}}^{\theta,\textrm{bdy}}[A_{x}^0,A_{y}^0]&=-\frac{1}{2}\int\frac{d\omega}{2\pi}A_j^0\,^*\left\{- \lim_{u=\epsilon\to 0} \omega^2\left[\mathbb{A}_{jk}+\mathbb{B}_{jl}\mathbb{O}_{l(i)}\left(\mathbb{S}^{-1}\right)_{(i)k} \right]\biggr|_{\textrm{on-shell}}^{\textrm{infalling}}\right\}A_k^0,
\label{SKR8}
\end{align}
where the indexes $j,k,l$ above run over $\left\{1\equiv x,2\equiv y\right\}$, while the index $(i)=1,2$ spans the different sets of initial conditions \eqref{ICs}. From the above result, one finally reads off the formal result for the thermal retarded Green's functions in the bulk magnetically condensed phase,
\begin{align}
G_{ij}^{(R),\textrm{cond}}(\theta,\omega)=- \lim_{u=\epsilon\to 0} \omega^2\left[\mathbb{A}_{jk}+\mathbb{B}_{jl}\mathbb{O}_{l(i)}\left(\mathbb{S}^{-1}\right)_{(i)k} \right]\biggr|_{\textrm{on-shell}}^{\textrm{infalling}}.
\label{condprop}
\end{align}

Then, from linear response theory, one has the following Kubo formulas for the diagonal and Hall conductivities in the magnetically condensed phase, respectively,
\begin{align}
\sigma_{xx}^{\textrm{cond}}(\theta,\omega)&=\sigma_{yy}^{\textrm{cond}}(\theta,\omega)= -\frac{G_{xx}^{(R),\textrm{cond}}(\theta,\omega)}{i\omega}=-i\omega\lim_{u=\epsilon\to 0} \mathbb{B}_{xx} \left[\mathbb{O}_{x(1)}\left(\mathbb{S}^{-1}\right)_{(1)x} + \mathbb{O}_{x(2)}\left(\mathbb{S}^{-1}\right)_{(2)x}\right]\biggr|_{\textrm{on-shell}}^{\textrm{infalling}}, \label{condiag}\\ \sigma_{xy}^{\textrm{cond}}(\theta,\omega)&=-\sigma_{yx}^{\textrm{cond}}(\theta,\omega)= -\frac{G_{xy}^{(R),\textrm{cond}}(\theta,\omega)}{i\omega}= -i\omega\lim_{u=\epsilon\to 0} \left\{\mathbb{A}_{xy} + \mathbb{B}_{xx} \left[\mathbb{O}_{x(1)}\left(\mathbb{S}^{-1}\right)_{(1)y} + \mathbb{O}_{x(2)}\left(\mathbb{S}^{-1}\right)_{(2)y}\right]\right\} \biggr|_{\textrm{on-shell}}^{\textrm{infalling}}. \label{condhall}
\end{align}

\subsubsection{Numerical results}
\label{sec:num}

The AC conductivities \eqref{condiag} and \eqref{condhall} generally have nontrivial real and imaginary parts, and they need to be numerically evaluated for different values of frequency $\omega$, $\theta$-angle, and mass profiles $m(u)$. For numerical calculations one needs to specify a background, and here I will work with the AdS$_4$-Schwarzschild metric given by Eq. \eqref{4.5} (as done in Ref. \cite{Rougemont:2015gia}). In order to express the results in terms of dimensionless quantities, I am going to work with the following dimensionless frequency (as also used in Ref. \cite{Rougemont:2015gia}),
\begin{align}
\mw:=\frac{3\omega}{4\pi T},
\label{eq:dimw}
\end{align}
and mass profiles given by,
\begin{align}
m(u)=\Lambda M(u),
\label{eq:mass}
\end{align}
where $\Lambda$ is the mass scale of the bulk monopole condensate \cite{Rougemont:2015gia} and I consider here numerical solutions for $M(u)=\tanh(u)$ and $M(u)=\tanh(u^2)$, with different values of the dimensionless combination $C_\Lambda\equiv\Lambda L/2=0,1,2$, as done in Ref. \cite{Rougemont:2015gia}.

The main steps involved in the numerical routine I developed are schematically as follows:
\begin{enumerate}[i.]
\item I substitute the AdS$_4$-Schwarzschild background \eqref{4.5} and some chosen profile for $M(u)$ in \eqref{eq:mass} back into the coupled ODE's \eqref{Ktxeom} and \eqref{Ktyeom}, which are then written in terms of the control parameters $(C_\Lambda,\theta,\mw)$;
\item I choose to work here with second order infrared expansions in \eqref{IRexp} and algebraically fix all the subleading infrared coefficients in terms of the leading ones, as discussed before;
\item Next I fix the two free leading infrared coefficients using the first set of initial conditions in Eq. \eqref{ICs}, specifying $\varphi_{1(1)}$ and $\varphi_{2(1)}$;
\item With this I calculate the numerical values of $K_{1(1)}$, $K'_{1(1)}$, $K_{2(1)}$, and $K'_{2(1)}$ truncated at second order close to the horizon, which are the required set of horizon conditions needed to initialize the numerical integration of the equations of motion;
\item Since the horizon and the boundary are singular points of the ODE's, I initialize the numerical integration slightly beyond the horizon, at $u_{\textrm{start}}=1-\epsilon$, and end the integration slightly below the boundary at the ultraviolet numerical cutoff $\epsilon=10^{-8}$;
\item I use the small value $\mw_{\textrm{start}}=10^{-5}$ as a proxy for the DC limit and run loops in $(\theta,\mw)$ with $\theta$ varying from 0 to 2 in steps of 0.1 and $\mw$ ranging from $\mw_{\textrm{start}}$ to 12 also in steps of 0.1 (clearly, other values may be chosen as a matter of convenience);
\item I repeat the previous step for $C_\Lambda=0,1,2$ and store the values of $\{C_\Lambda,\theta,\mw,K_{1(1)}(\epsilon),K'_{1(1)}(\epsilon),K_{2(1)}(\epsilon),K'_{2(1)}(\epsilon)\}$;
\item Next I repeat items iii to vii, but now for the second set of initial conditions in Eq. \eqref{ICs}, and store the values of $\{C_\Lambda,\theta,\mw,K_{1(2)}(\epsilon),K'_{1(2)}(\epsilon),K_{2(2)}(\epsilon),K'_{2(2)}(\epsilon)\}$;
\item From the previous results, for each value of $C_\Lambda$, I construct the matrices $\mathbb{S}_{a(i)}=K_{a(i)}(\epsilon)$ and $\mathbb{O}_{a(i)}=K'_{a(i)}(\epsilon)$ (there will be one of such matrices for each value of the control parameters $(\theta,\mw)$), and also invert all the generated matrices $\mathbb{S}_{a(i)}$ to obtain $\left(\mathbb{S}^{-1}\right)_{(i)a}$;
\item With all these results at hand, I finally calculate the real and imaginary parts of the diagonal and Hall conductivities for each value of the control parameters $(C_\Lambda,\theta,\mw)$ using the holographic Kubo formulas \eqref{condiag} and \eqref{condhall}.  
\end{enumerate}

The numerical robustness and accuracy of the method implemented above has been nontrivially checked in many different ways. First, the numerical results obtained at $C_\Lambda=0$ (corresponding to no monopoles in the bulk) coincide with the analytical results of the Maxwell theory with the $\theta$-term reviewed in section \ref{sec:diluted}, as it should be. Second, for $C_\Lambda\neq 0$ (corresponding to the magnetically condensed phase) the results for the real and imaginary parts of the diagonal and Hall conductivities converge to the analytical results of the Maxwell theory in the ultraviolet regime of large frequencies, as expected. Moreover, for $\theta=0$ I was able to reobtain the numerical results of Ref. \cite{Rougemont:2015gia}, which were computed using a completely different method based on first order flow equations.

The results for the real and imaginary parts of the diagonal and Hall conductivities for different effective mass profiles $m(u)$ are displayed in Figs. \ref{fig1} -- \ref{fig8}.

The main conclusion of the present work, drawn from the analysis of these plots, is that the real and imaginary parts of the diagonal and Hall conductivities in all the cases vanish in the DC limit ($\mw\to 0$) when there is a magnetic monopole condensate in the bulk ($C_\Lambda\neq 0$). This generalizes the conclusion of Ref. \cite{Rougemont:2015gia}, showing that not only the diagonal, but also the Hall DC conductivity in the strongly coupled QFT vanishes as a consequence of the presence of a magnetic monopole condensate in the bulk. Notice also that the inclusion of the topological $\theta$-term does not change this conclusion for the DC limit of the diagonal conductivity, even though at finite frequencies the results are sensitive to the value of the $\theta$-angle. Therefore, we see that a magnetic monopole condensate in the bulk provides a fairly general and robust mechanism for generating strongly coupled QFT's with vanishing DC conductivities.

The regime of intermediate frequencies also unveils some interesting features. Notice from the pairs of Figs. \ref{fig1} and \ref{fig2}, \ref{fig3} and \ref{fig4}, \ref{fig5} and \ref{fig6}, \ref{fig7} and \ref{fig8}, that for a fixed effective mass profile $m(u)=\Lambda M(u)$, around the regions where the oscillation amplitudes of the AC conductivities are larger, the real part of the Hall conductivity is pretty similar to the imaginary part of the diagonal conductivity, while the real part of the diagonal conductivity is very similar to minus the imaginary part of the Hall conductivity, even though they clearly differ in the ultraviolet limit of large frequencies.

Moreover, by looking at the pairs of Figs. \ref{fig1} and \ref{fig3}, \ref{fig2} and \ref{fig4}, \ref{fig5} and \ref{fig7}, \ref{fig6} and \ref{fig8}, one also notices that the regions where the oscillation amplitudes of the AC conductivities are larger tend to be shifted toward larger values of the frequency for increasing values of the characteristic mass scale of the monopole condensate. This is more clearly illustrated in Figs. \ref{fig9} -- \ref{fig12}, where the conductivities are plotted as functions of the monopole condensate scale and the frequency.

\section{Conclusion}
\label{sec:conc}

In this work I investigated the effects caused by the topological $\theta$-angle in the diagonal and Hall AC conductivities of strongly coupled 3-dimensional QFT's holographically dual to a 4-dimensional bulk condensate of magnetic monopoles. In this way, I generalized the work of Ref. \cite{Rougemont:2015gia}, whose results constitute a particular case of the present work with $\theta=0$. This generalization not only allowed for the numerical calculation of nontrivial profiles for the AC Hall conductivity, but it also uncovered how the diagonal AC conductivity is affected by the $\theta$-angle.

The main conclusion of the present work regards the infrared limit of zero frequencies. In fact, it was concluded that a monopole condensate in the bulk constitutes a fairly general and robust holographic mechanism to generate dual strongly coupled QFT's with vanishing DC diagonal and Hall conductivities.

In the opposite, ultraviolet limit of large frequencies, the diagonal and Hall conductivities converge to the analytical results of the Maxwell theory. This can be physically traced back to the fact that while perturbations of very low frequencies are sensitive to the magnetically condensed phase in the bulk, very high frequency disturbances are not. In fact, in the DC limit of zero frequencies the bulk monopole condensate can be effectively seen as a macroscopically continuous dual superconducting medium, and the implied confinement of electric flux tubes within the bulk makes the transport of electric charges -- seen as the intersection of these bulk flux tubes with the boundary, as discussed in Refs. \cite{faulkner-iqbal,iqbal} -- negligible, therefore leading to a vanishing DC conductivity. On the other hand, since very high frequency perturbations can microscopically resolve the structure of the condensate of magnetic monopoles and probe distances much smaller than the characteristic distance scale of the magnetic condensate, $\Lambda^{-1}$, charge transport at the boundary in this ultraviolet limit takes place as in the diluted (Maxwell) phase.

In between, for intermediate frequencies, the interplay between these frequencies, the $\theta$-angle, and the characteristic mass scale of the monopole condensate $\Lambda$ unveiled strong correlations between the profiles for the real part of the Hall the conductivity and the imaginary part of the diagonal conductivity, and also between the profiles for the real part of the diagonal conductivity and (minus) the imaginary part of the Hall conductivity, specially within the region of parameters where the oscillation amplitudes of the AC conductivities are larger.

I worked here in the probe approximation with a fixed black hole background, and I intend to generalize in the future the calculations pursued here by considering the backreaction of the matter action describing the monopole condensate into a dynamical background.

\acknowledgments

I thank Jorge Noronha for interesting questions about this manuscript. I also acknowledge financial support by Universidade do Estado do Rio de Janeiro (UERJ) and Funda\c{c}\~{a}o Carlos Chagas de Amparo \`{a} Pesquisa do Estado do Rio de Janeiro (FAPERJ).

\appendix
\section{Hall and diagonal conductivities in the bulk electrically condensed phase}
\label{sec:eltcond}

In this appendix I discuss the case with a condensate of electric charges in the bulk. The case with no $\theta$-term was originally presented in Ref. \cite{Rougemont:2015gia} and gives qualitatively the same results obtained for the holographic superconductor proposed in Ref. \cite{Hartnoll:2008vx}.

As detailed discussed in Ref. \cite{Rougemont:2015gia}, in the electrically condensed phase the effective action for the vector field sector is the Proca action with a radial-dependent mass which vanishes at the boundary (where the massive Proca field reduces to the massless Maxwell field sourcing a conserved vector current operator in the dual QFT). In an ultraviolet completion of such action the radial-dependent mass, which I take as a prescribed profile here, corresponds to a dynamical scalar field associated to the electric condensate (the potential of this scalar field may be chosen such as to produce some prescribed profile used for the radial-dependent mass in the effective Proca action). By including a $\theta$-term, the low energy effective action describing the lowest-lying modes of the electrically condensed phase in the bulk reads,
\begin{align}
S_{\textrm{cond}}^{\theta,\textrm{elt}}[A_\mu]&=-\int_{\mathcal{M}_4}d^4x\sqrt{-g}\left[\frac{1}{4}F_{\mu\nu}^2 +\frac{m^2(u)}{2}A_\mu^2\right]-\frac{\theta}{4}\int_{\mathcal{M}_4}d^4x\sqrt{-g} F_{\mu\nu}\tilde{F}^{\mu\nu}\nonumber\\
&=-\frac{1}{2}\int_{\partial\mathcal{M}_4}d^3x \left(\sqrt{-g}g^{uu}g^{\nu\beta} A_\nu F_{u\beta} +\theta\varepsilon^{u\nu\alpha\beta}A_\nu\partial_\alpha A_\beta\right)\Biggr|_{u=\epsilon}^{u=u_H} +\frac{1}{2}\int_{\mathcal{M}_4}d^4x A_\nu \left[\partial_\mu\left(\sqrt{-g}g^{\mu\alpha}g^{\nu\beta} F_{\alpha\beta}\right)+\right.\nonumber\\
&\left. -\sqrt{-g}m^2(u)g^{\mu\nu}A_\mu\right],
\label{SProca1}
\end{align}
where one sees by comparing the above action with Eq. \eqref{3.3} that the \emph{off-shell} border terms have the same functional form as in the diluted phase described by Maxwell theory\footnote{Without fixing the radial gauge in Maxwell theory.} with the $\theta$-term, but now the equation of motion is given by Proca equation,
\begin{align}
\partial_\mu\left(\sqrt{-g}g^{\mu\alpha}g^{\nu\beta} F_{\alpha\beta}\right)-\sqrt{-g}m^2(u)g^{\mu\nu}A_\mu=0.
\label{ProcaEOM1}
\end{align}
The $\nu=x(y)$ component of Proca equation \eqref{ProcaEOM1} reads in momentum space,\footnote{I consider again the limit of zero spatial momentum, which is enough for the calculation of the conductivities.}
\begin{align}
\partial_u\left(\sqrt{\frac{g_{tt}}{g_{uu}}}A'_{x(y)}\right)+\sqrt{\frac{g_{uu}}{g_{tt}}} \left[\omega^2-m^2(u)g_{tt}\right] A_{x(y)}=0,
\label{ProcaEOM2}
\end{align}
while the $\nu=u$ component gives a constraint expressing $A_u$ as function of $A'_t$, and the remaining equation, after using the constraint for $A_u$, gives a decoupled equation of motion for $A_t$. Since the relevant components of the vector field for the calculation of the conductivities, $A_x$ and $A_y$, satisfy the same decoupled equation of motion \eqref{ProcaEOM2}, which is also the same one obtained in Ref. \cite{Rougemont:2015gia} for the Proca theory without the $\theta$-term, the numerical solutions are also the same which were derived in Ref. \cite{Rougemont:2015gia}. Moreover, as $A_u$ and $A_t$ do not couple to $A_x$ and $A_y$ (neither in the equations of motion, nor in the boundary action), the relevant sector of the boundary action for the calculation of the diagonal and Hall conductivities has the same \emph{off-shell} functional form of Eq. \eqref{Smom2},
\begin{align}
\mathcal{S}_{\textrm{cond}}^{\theta,\textrm{elt,bdy}}[A^0_x,A^0_y]&=-\frac{1}{2}\int \frac{d\omega}{2\pi} \left\{-\lim_{u=\epsilon\to 0}\left[ \sqrt{-g} g^{uu} g^{xx}\left(A_x^*A_x'+A_y^*A_y'\right) +i\omega\theta\left(A_x^*A_y-A_y^*A_x\right)\right]\right\} \Biggr|_{\textrm{on-shell}}^{\textrm{infalling}}.
\label{SProca2}
\end{align}
But one notes that the \emph{on-shell} action \eqref{SProca2} will be actually different from Eq. \eqref{Smom3} (valid for the diluted Maxwell phase), because the on-shell components of the Proca field solve Eq. \eqref{ProcaEOM2} instead of Maxwell equations \eqref{3.10} or \eqref{3.11}.

Since $m(u=0)=0$, one recovers in the ultraviolet limit of high frequencies the results for the Maxwell theory with the $\theta$-term revised in section \ref{sec:diluted}. In particular, since the terms proportional to $\theta$ in the on-shell boundary action \eqref{SProca2} do not depend on the radial derivative of $A_{x(y)}$, the AC Hall conductivities in the electrically condensed phase are exactly the same ones obtained for the diluted phase in Eq. \eqref{dilcond}. On the other hand, the diagonal AC conductivities in the electrically condensed phase are given by the same numerical results derived in Ref. \cite{Rougemont:2015gia} for the Proca theory without the $\theta$-term. In this way, the addition of the $\theta$-term to the Proca theory does not modify the diagonal conductivities of the latter, which therefore still diverge in the DC limit of zero frequency \cite{Rougemont:2015gia}, indicating a superconducting state at the boundary QFT in the same lines of Ref. \cite{Hartnoll:2008vx}, while such addition does provide finite and constant Hall conductivities which coincide with the result for Maxwell theory with the $\theta$-term \cite{Iqbal:2008by}. Strongly coupled holographic superconducting states with nonzero Hall conductivity have been previously considered, for instance, in Refs. \cite{Chen:2011ny,remarkshall}.

\end{document}